\documentclass[aps,physrev,reprint,groupedaddress]{revtex4-2}

\usepackage{graphicx}
\usepackage{subcaption}
\usepackage{amsmath, amssymb, amsfonts}
\usepackage{amsthm}
\usepackage{physics}
\usepackage{xcolor}
\usepackage{hyperref}
\hypersetup{colorlinks,allcolors=blue}
\theoremstyle{definition}
\newtheorem{definition}{Definition}
\theoremstyle{definition}
\newtheorem{protocol}{Protocol}
\theoremstyle{definition}
\newtheorem*{observation}{Observation}
\theoremstyle{plain}
\newtheorem{proposition}{Proposition}

\begin{document}


\title{Coherence transfer under bounded channel capacity}


\author{Dhruv Baheti}
\affiliation{Institute of Informatics, National Quantum Information Centre, Faculty of Mathematics, Physics and Informatics, University of Gda\'nsk, Wita Stwosza 57, 80-308 Gda\'nsk, Poland}

\author{Karol Horodecki}
\affiliation{Institute of Informatics, National Quantum Information Centre, Faculty of Mathematics, Physics and Informatics, University of Gda\'nsk, Wita Stwosza 57, 80-308 Gda\'nsk, Poland}


\date{\today}

\begin{abstract}
We investigate quantum coherence transfer between two parties whose local reference bases are misaligned, under a constraint on the capacity of the quantum communication channel. Within the resource theory of coherence, we formulate the task as maximizing the coherence received by Bob when Alice is restricted to Maximally Incoherent Operations (MIO) in her local basis. We first analyze two single-shot protocols, based on direct and locally entangled transmission, and derive their phase-optimized coherence yields. We further investigate their performance when Bob's reference basis is unknown, averaging over uniformly distributed basis configurations. In the asymptotic regime, we exploit the reversibility of coherence under MIO to characterize achievable transfer rates in terms of Alice's initial coherence. For single-qubit transmission, we derive an analytical lower bound using pure-state conversions and identify a regime in which the optimal yield saturates the channel capacity.
\end{abstract}


\maketitle


The departure of quantum mechanics from classical physics is fundamentally rooted in the principle of superposition. When manifested in multipartite systems, this feature gives rise to quantum entanglement, a central property of quantum mechanics that has historically played a significant role in quantum information science \cite{Horodecki2009_Entanglement}. However, even within a single quantum system, superposition gives rise to another foundational quantum feature: quantum coherence. Closely related to entanglement, quantum coherence characterizes superposition relative to a chosen reference basis and underlies the interference phenomena exhibited by quantum systems. While the study of coherence originates from the field of quantum optics \cite{Glauber1963, MandelWolf1995}, it has more recently been formalized and studied within the framework of quantum information theory \cite{Baumgratz2014_Coherence, Streltsov2017_CoherenceReview, Levi_2014, aberg2006}.

A distinguishing feature of quantum coherence is its dependence on the choice of reference basis \cite{speakable_unspeakable, Gour_2008}. Unlike entanglement, which is invariant under local unitary transformations, the coherence attributed to a quantum state generally changes with the reference basis in which it is evaluated \cite{Streltsov2017_CoherenceReview, Bartlett2007_ReferenceFrames}. This distinction becomes particularly relevant in quantum communication between spatially separated parties, where the sender and receiver may not possess a shared reference frame. The challenges posed by reference-frame misalignment have been studied extensively in quantum information processing, including the development of reference-frame-independent communication and cryptographic protocols \cite{Bartlett2007_ReferenceFrames, Laing2010_RFI, DAmbrosio2012_FrameDrift, experiment_qc}. From the perspective of coherence, such scenarios raise a related but distinct question: \textit{how much coherence, as quantified in the receiver's local reference basis, can be transferred from a sender operating under a different coherence theory?}

Addressing this question requires accounting not only for the relative orientation of the parties' reference bases, but also for the operations available to them. The resource-theoretic framework of coherence provides a natural setting for this investigation, with different classes of incoherent operations characterizing the transformations accessible to a locally constrained agent \cite{RT_Coherence, Baumgratz2014_Coherence, ChitambarGour2016, ChitambarGour2016_Comparison, distributed_coherence}. In addition, the established connections between coherence and entanglement suggest potential strategies for manipulating the coherence properties of transmitted states. In particular, local coherence can be converted into bipartite entanglement through incoherent operations, while the resulting entanglement remains invariant under local changes of basis \cite{Streltsov2015, Killoran2016_ConvertingCoherence, ChitambarHsieh2016, Ma2017, Sen2021}. These observations motivate the two coherence-transfer schemes considered in this work: one based on \textit{direct transmission} of locally prepared states, and another employing an \textit{entangled encoding} before transmission. By comparing these approaches, we investigate how the choice of encoding and knowledge of the receiver's reference basis influence the coherence delivered between agents with distinct local coherence theories.

\begin{figure}[htbp]
    \centering
    \includegraphics[width=\linewidth]{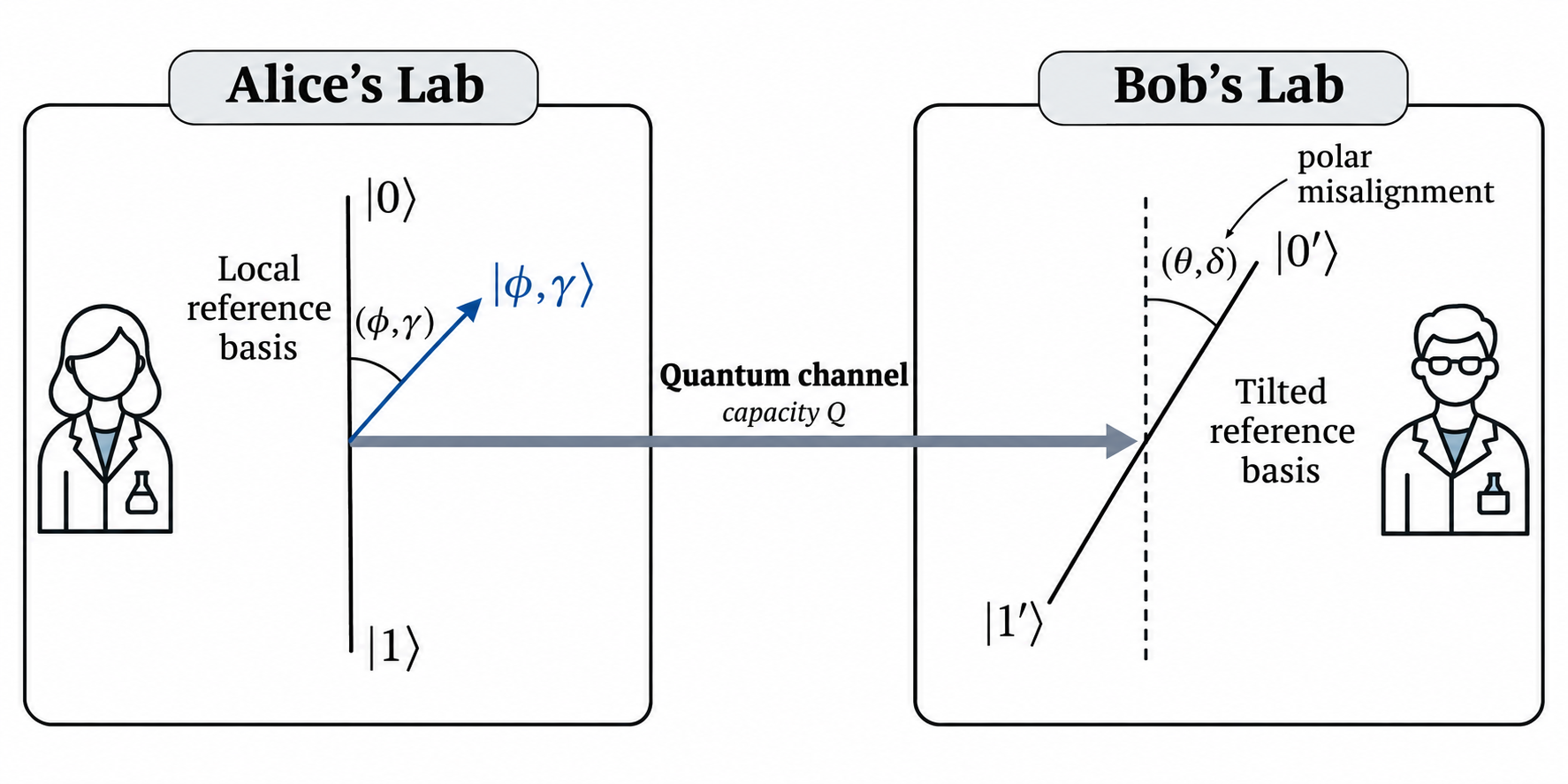}
    \caption{Schematic of the coherence transfer scenario. Alice prepares a coherent quantum state and transmits it to Bob through a quantum channel of bounded capacity $Q$. At a distant location, Bob performs measurements in a rotated local reference basis.}
    \label{fig:scenario_reference_frames}
\end{figure}

In this work, we formulate a coherence-transfer scenario between two agents equipped with distinct local coherence theories and connected by an ideal quantum communication channel of bounded capacity. The capacity constraint is essential to obtaining a meaningful optimization problem, since unrestricted communication would allow the sender to prepare and transmit arbitrarily many states that are incoherent in her own reference basis but coherent in the receiver's. Within this framework, we investigate two explicit single-shot transfer schemes, analyzing their performance both when the sender knows the receiver's reference basis and when that basis is unknown. We derive analytical conditions for optimizing the delivered coherence and compare the expected yields under uniformly distributed reference-frame misalignments. We further extend the analysis to the asymptotic regime, where the reversibility of coherence under maximally incoherent operations enables state conversion protocols and yields an analytical lower bound on the optimal coherence transfer, together with conditions for saturating the communication capacity.

\section{Operational Setting and Definitions}
\label{sec:scenario}

In this section, we establish a general framework for coherence transfer between parties with misaligned local reference bases, subject to a bounded quantum communication capacity.

Consider two agents, Alice and Bob, each equipped with a resource theory of coherence on a $d$-dimensional Hilbert space $\mathcal{H}_d$. Their respective incoherent bases are denoted by
$\mathcal{B}_A=\{\ket{i_A}\}_{i=1}^{d}$ and
$\mathcal{B}_B=\{\ket{i_B}\}_{i=1}^{d}$.
For $X\in\{A,B\}$, the corresponding set of free states is
\begin{equation}
    \mathcal{I}_X
    =
    \left\{
    \sum_{i=1}^{d}p_i\ket{i_X}\bra{i_X}
    \,\middle|\,
    p_i\geq 0,\ \sum_i p_i=1
    \right\}.
\end{equation}
The two reference bases are related by a unitary transformation $U$, such that $\ket{i_B}=U\ket{i_A}$.

Both parties quantify coherence using the relative entropy of coherence \cite{Baumgratz2014_Coherence, Streltsov2017_CoherenceReview},
\begin{equation}
    C_X(\rho)=S(\Delta_X(\rho))-S(\rho),
\end{equation}
where $S$ denotes the von Neumann entropy and $\Delta_X$ is the completely dephasing channel in the reference basis of agent $X$, defined as
\begin{equation}
    \Delta_X(\rho)
    =
    \sum_{i=1}^{d}
    \ket{i_X}\bra{i_X}\rho\ket{i_X}\bra{i_X}.
\end{equation}
For a collection of $r$ elementary systems, the local resource theory is defined relative to the product basis
$\mathcal{B}_X^{\otimes r}$, with associated incoherent-state set $\mathcal{I}_X^{(r)}$ consisting of all density operators diagonal in this basis.

We now introduce the communication setting. Alice initially possesses $m$ copies of a state $\rho\in\mathcal{S}(\mathcal{H}_d)$, with local coherence $C_A(\rho)$.  Alice and Bob are connected by ideal noiseless quantum communication channel with quantum capacity $Q$ qubits per channel use. Accordingly, $n$ channel uses allow Alice to transmit a quantum register of dimension at most $2^{nQ}$. The effective channel is given as $\operatorname{id}_{2^{nQ}}^{A\to B}$. Our objective is to determine:
\begin{quote}
    \textit{Subject to these constraints, how much coherence can Alice deliver to Bob, measured in Bob's local reference basis, per use of the quantum channel?}
\end{quote}

To investigate this question, we allow Alice to perform arbitrary maximally incoherent operations (MIO), the largest class of quantum channels preserving the set of incoherent states. For input and output systems consisting of $m$ and $m'$ elementary subsystems, respectively, MIO consists of all completely positive trace-preserving (CPTP) maps satisfying
\begin{equation}
    \mathrm{MIO}_X^{m\to m'}
    =
    \left\{
    \Lambda
    \,\middle|\,
    \Lambda(\mathcal{I}_X^{(m)})
    \subseteq
    \mathcal{I}_X^{(m')}
    \right\}.
\end{equation}

\begin{definition}[Admissible Coherence-Transfer Protocol]
\label{def:admissible_protocol}
For an input of $m$ copies of a $d$-dimensional state and $n$ uses of a quantum channel with capacity $Q$, an admissible $(m,n,Q)$ coherence-transfer protocol is a map
\begin{equation}
    \mathcal{P}_Q^{n,m}:
    \mathcal{S}(\mathcal{H}_d^{\otimes m})
    \longrightarrow
    \mathcal{S}(\mathcal{H}_d^{\otimes m'}),
\end{equation}
satisfying
\begin{equation}
    \mathcal{P}_Q^{n,m}\in\mathrm{MIO}_A^{m\to m'},
    \qquad
    d^{m'}\leq 2^{nQ}.
\end{equation}
The set of all admissible $(m,n,Q)$ protocols is denoted by $\mathfrak{P}_Q^{n,m}$. The coherence-transfer yield of a protocol $\mathcal{P}_Q^{n,m}$ is defined as the amount of coherence delivered to Bob per channel use:
\begin{equation}
    \mathcal{Y}_{\mathcal{P}_Q^{n,m}}(\rho)
    :=
    \frac{1}{n}
    C_B\left(
    \mathcal{P}_Q^{n,m}(\rho^{\otimes m})
    \right).
\end{equation}
\end{definition}

We now motivate the restriction to finite quantum channel capacity for distinct local theories, i.e., $\mathcal{I}_A\neq\mathcal{I}_B$. Suppose that $Q$ is allowed to grow without bound. Because the local incoherent-state sets are distinct, there exists at least one Alice-incoherent basis state $\ket{i_A}$ for which
\begin{equation}
    C_B(\ket{i_A}\bra{i_A})=c>0.
\end{equation}
Alice may prepare this state freely using the replacement channel
\begin{equation}
    \mathcal{R}_d(\omega)
    =
    \operatorname{Tr}[\omega]\ket{i_A}\bra{i_A},
\end{equation}
which belongs to $\mathrm{MIO}_A$. For a fixed number of channel uses $n$, choose $m'$ such that $d^{m'} \approx 2^{nQ}$.
Alice can then prepare and transmit $m'$ copies of $\ket{i_A}$, independently of her initial resource state. The resulting coherence-transfer yield is
\begin{equation}
    \mathcal{Y}_{\mathcal{P}_Q^{n,m}}(\rho)
    =
    \frac{c\,m'}{n}
    \approx \frac{c\,Q}{\log d}
    \xrightarrow[Q\to\infty]{}\infty.
\end{equation}

This result is not surprising by any means, it is a definitional consequence of resource theories. Resource theories constrain an agent's capacity to manipulate a quantum state. Unrestricted access to quantum communication with an unconstrained or differently constrained agent bypasses the receiving agents operational restrictions. However, an interesting and non-trivial range of theories with $0 < Q < \infty$ emerges, where an agent may obtain some external assistance while still being meaningfully constrained by their local resource theory.

\begin{definition}[Optimal Coherence-Transfer Yield]
\label{def:optimal_yield}
The optimal coherence-transfer yield is the supremum of the coherence-transfer yield over all admissible protocols:
\begin{equation}
    \widehat{\mathcal{Y}}_Q^{n,m}(\rho)
    :=
    \sup_{\mathcal{P}_Q^{n,m}\in\mathfrak{P}_Q^{n,m}}
    \mathcal{Y}_{\mathcal{P}_Q^{n,m}}(\rho).
\end{equation}
\end{definition}

Since the delivered coherence cannot exceed the logarithm of the transmitted Hilbert-space dimension, we obtain
\begin{equation}
    0\leq\widehat{\mathcal{Y}}_Q^{n,m}(\rho)\leq Q.
\end{equation}
For distinct local incoherent-state sets, the lower bound is strict whenever $nQ\geq\log d$, since Alice can freely prepare and transmit a state that is incoherent in her basis but coherent in Bob's. While the optimal coherence-transfer yield characterizes the best performance under the prescribed constraints, its evaluation requires optimization over the full class of admissible MIO protocols. In the following section, we instead investigate two explicit single-shot coherence-transfer schemes with pure state inputs and compare their achievable yields under different assumptions about Alice's knowledge of Bob's reference basis.

\section{Single Shot Coherence Transfers}
\label{sec:scenario_raw_yields}

In this section, we consider a scenario where the sender Alice is constrained by a qubit coherence theory with the reference basis $\mathcal{B}_A = \{|0\rangle, |1\rangle\}$. Bob's local measurement basis is rotated with respect to Alice's reference frame by a polar misalignment angle $\theta \in [0, \frac \pi 2]$ and a phase misalignment angle $\delta \in [0, 2\pi)$. Consequently, Bob's local measurement basis $\mathcal{B}_B = \{|\overline{0}\rangle, |\overline{1}\rangle\}$ is expressed in Alice's frame as:
\begin{align}
|\overline{0}\rangle &= \cos\theta |0\rangle + e^{i\delta} \sin\theta |1\rangle, \label{eq:bob_basis_0} \\
|\overline{1}\rangle &= -e^{-i\delta} \sin\theta |0\rangle + \cos\theta |1\rangle. \label{eq:bob_basis_1}
\end{align}
We look at a class of scenarios corresponding to the tuple $(m,n,Q) = (1,1,2)$. Alice possesses a single ($m = 1$) pure state parametrized by the initial tilt angle $\phi \in [0, \frac \pi 2]$ and local phase $\gamma \in [0, 2\pi)$:
\begin{equation}
|\phi, \gamma\rangle = \cos\phi |0\rangle + e^{i\gamma} \sin\phi |1\rangle.
\end{equation}

\begin{figure}[htbp]
    \centering
    \includegraphics[width=\linewidth]{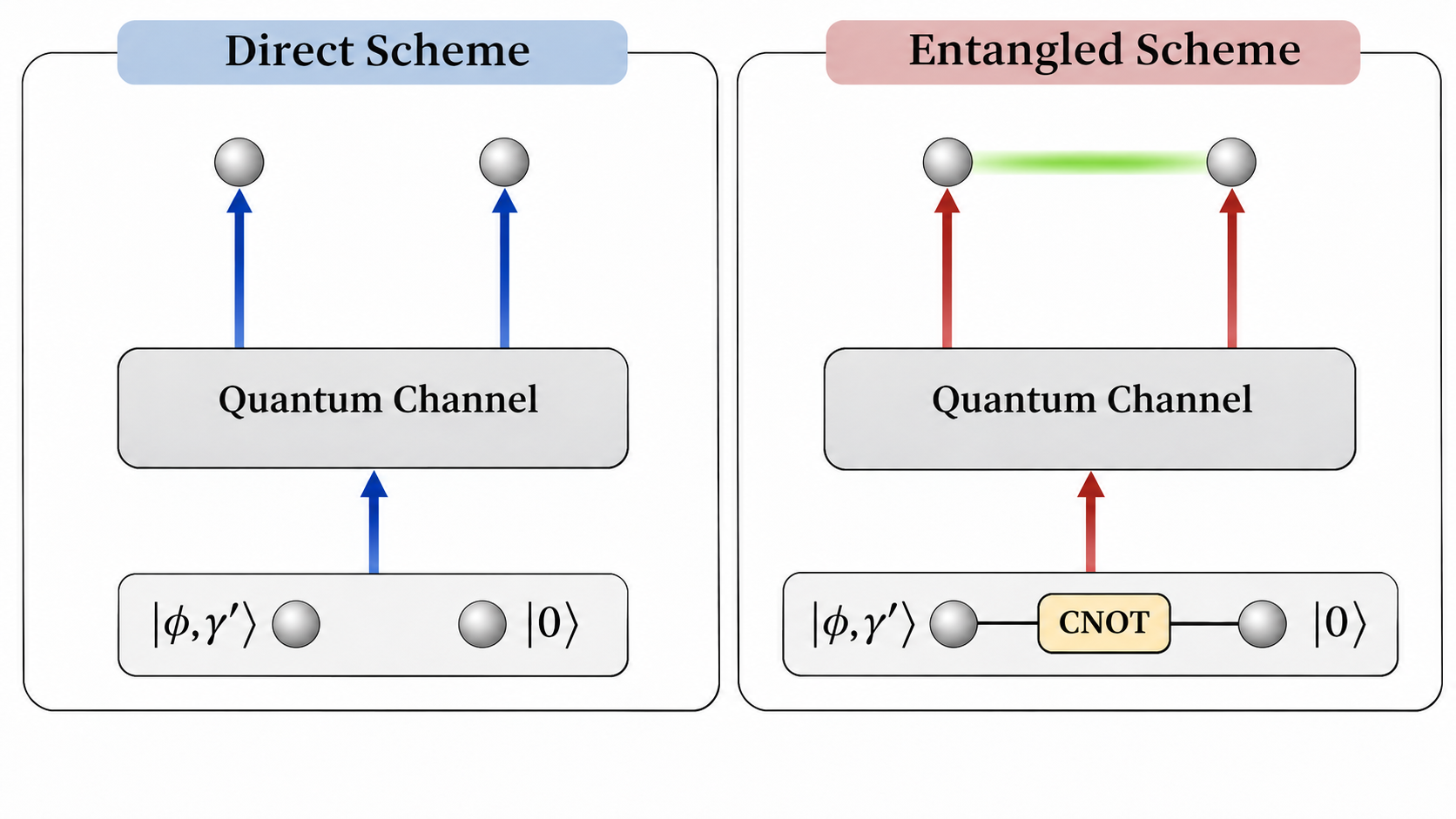}
    \caption{ Schematic of the two single-shot coherence-transfer schemes.}
    \label{fig:scheme_comparison}
\end{figure}

The parties are permitted a single use $(n = 1)$ of a two-qubit channel ($Q = 2$). We consider two coherence-transfer protocols within this setting: the \emph{Direct Scheme} and the \emph{Entangled Scheme}. In the former, Alice attaches an incoherent ancilla to her input state and transmits the resulting product state, while in the latter, she first applies an incoherent controlled-NOT operation to generate an entangled state before transmission. In both protocols, Alice may freely adjust the phase of her input using incoherent phase operations. We describe these constructions and derive their respective coherence yields in the following subsections.

\smallskip

Since we restrict our attention to the $(m,n,Q)=(1,1,2)$ setting throughout this section, we henceforth drop these indices and denote the two protocols simply by $\mathcal{P}^{\mathrm{dir}}$ and $\mathcal{P}^{\mathrm{ent}}$ for notational brevity.

\subsection{The Direct Scheme}

\begin{protocol}[Direct Scheme]
\label{prot:direct}
Alice appends an incoherent ancilla $\ket{0}$ to her input state
$\ket{\phi,\gamma}$ and applies an incoherent phase operation
$\gamma\mapsto\gamma'$. The resulting product state is transmitted
to Bob:
\begin{equation}
    \mathcal{P}^{\mathrm{dir}}(\ket{\phi,\gamma})
    =
    \ket{\phi,\gamma'}\otimes\ket{0}.
\end{equation}
\end{protocol}

The motivation for this construction is its geometric complementarity. For an input with nonzero coherence in Alice's basis, the states $\ket{\phi,\gamma'}$ and $\ket{0}$ are oriented along distinct axes on the Bloch sphere and hence cannot both be incoherent in any single orthonormal basis. Thus, the product encoding guarantees a strictly positive delivered coherence yield, independently of the relative orientation of the two reference bases.

To quantify this yield, we consider the outcome probabilities associated with Bob's local basis $\{\ket{\overline{0}},\ket{\overline{1}}\}$. For the tilted qubit, the probability of obtaining $\ket{\overline{0}}$ is
\begin{equation}
    \begin{aligned}
    p
    &= |\langle \overline{0} | \phi, \gamma' \rangle|^2 \\
    &= \cos^2(\phi-\theta)
    - \sin(2\phi)\sin(2\theta)\sin^2\Omega,
    \end{aligned}
\end{equation}
where $\Omega=(\gamma'-\delta)/2$ is the effective relative phase. Since the transmitted state is pure, its relative entropy of coherence is given by the Shannon entropy of the corresponding incoherent-basis probability distribution $C_B(\ket{\psi}\bra{\psi}) = S(\Delta_B(\ket{\psi}\bra{\psi}))$. The tilted qubit therefore contributes $h(p)$, where $h(p)=-p\log_2 p-(1-p)\log_2(1-p)$ denotes the binary entropy.

For the incoherent ancilla $\ket{0}$, the corresponding outcome probabilities are $\cos^2\theta$ and $\sin^2\theta$, yielding an additional coherence contribution $h(\cos^2\theta)$. By additivity of the relative entropy of coherence on product states \cite{Baumgratz2014_Coherence}, the total delivered yield of the Direct Scheme is
\begin{multline}
    \mathcal{Y}_{\mathcal{P}^{\mathrm{dir}}}
    (\phi,\theta,\Omega)
    =
    C_B\left(
    \mathcal{P}^{\mathrm{dir}}(\ket{\phi,\gamma})
    \right)
    \\
    =
    h\Big(
    \cos^2(\phi-\theta)
    -
    \sin(2\phi)\sin(2\theta)\sin^2\Omega
    \Big)
    +
    h(\cos^2\theta).
    \label{eq:raw_yield_direct}
\end{multline}

\subsection{The Entangled Scheme}
\begin{protocol}[Entangled Scheme]
\label{prot:entangled}
Alice appends an incoherent ancilla $\ket{0}$ to her input state $\ket{\phi,\gamma}$ and applies an incoherent phase operation $\gamma\mapsto\gamma'$, followed by a controlled-NOT (CNOT) gate. The resulting entangled state is transmitted to Bob:
\begin{equation}
\begin{aligned}
    \mathcal{P}^{\mathrm{ent}}(\ket{\phi,\gamma})
    &=
    \mathrm{CNOT}
    \left(\ket{\phi,\gamma'}\otimes\ket{0}\right)\\
    &=
    \cos\phi\ket{00}
    +
    e^{i\gamma'}\sin\phi\ket{11}.
\end{aligned}
\end{equation}
\end{protocol}

The Entangled Scheme is motivated by two established connections between the resource theories of coherence and entanglement. First, the relative entropy of coherence can be converted into an equal amount of relative entropy of entanglement by applying a generalized CNOT operation to a coherent state and a sufficiently large incoherent ancilla \cite{Streltsov2015,Killoran2016_ConvertingCoherence, Ma2017}. Second, for any bipartite state $\rho$, the relative entropy of coherence in a local product basis satisfies $C_R(\rho)\geq E_R(\rho)$, where $E_R$ denotes the relative entropy of entanglement. This follows from the inclusion of incoherent states within the set of separable states \cite{Streltsov2017_CoherenceReview}.

\begin{observation}[Coherence Guarantee of the Entangled Scheme]
\label{obs:entangled_guarantee}
The Entangled Scheme guarantees a coherence-transfer yield at least as large as Alice's initial coherence, independently of the relative orientation of the two reference bases:
\begin{equation}
\begin{aligned}
    \mathcal{Y}_{\mathcal{P}^{\mathrm{ent}}} (\phi,\theta,\Omega) = C_B(\ket{\Psi_{\mathrm{ent}}})
    &\geq E_R(\ket{\Psi_{\mathrm{ent}}})\\
    &= C_A(\ket{\phi,\gamma}),
\end{aligned}
\end{equation}
where $\ket{\Psi_{\mathrm{ent}}} =\mathcal{P}^{\mathrm{ent}}(\ket{\phi,\gamma})$. Indeed, the transmitted pure state has Schmidt coefficients $\cos\phi$ and $\sin\phi$, giving $E_R(\ket{\Psi_{\mathrm{ent}}}) =h(\cos^2\phi)=C_A(\ket{\phi,\gamma})$. The inequality follows from \cite{Streltsov2017_CoherenceReview}. Consequently, the delivered yield is strictly positive for any coherent input state.
\end{observation}

The two-qubit state is transmitted to Bob, who evaluates its coherence in the product basis $\mathcal{B}_B^{\otimes 2} =\{\ket{\overline{ij}}\}_{i,j=0}^{1}$. The corresponding probabilities are
\begin{equation}
    p_{ij}(\theta, \phi, \Omega) = |\langle \overline{ij} | \Psi_{\text{ent}} \rangle|^2.
\end{equation}
These probabilities depend on the effective relative phase $\Omega=\gamma'-2\delta$ and are given by
\begin{align}
    p_{00}
    &=
    \cos^2\phi\cos^4\theta
    +
    \sin^2\phi\sin^4\theta
    \nonumber\\
    &\quad+
    \frac{1}{4}
    \sin(2\phi)\sin^2(2\theta)\cos\Omega,
    \nonumber\\[4pt]
    p_{11}
    &=
    \sin^2\phi\cos^4\theta
    +
    \cos^2\phi\sin^4\theta
    \nonumber\\
    &\quad+
    \frac{1}{4}
    \sin(2\phi)\sin^2(2\theta)\cos\Omega,
    \nonumber\\[4pt]
    p_{01}=p_{10}
    &=
    \frac{1}{4}\sin^2(2\theta)
    \left(1-\sin(2\phi)\cos\Omega\right).
    \label{ent_probs}
\end{align}

Since the transmitted state is pure, its relative entropy of coherence is the Shannon entropy of the incoherent-basis probability distribution. The total delivered yield of the Entangled Scheme is therefore
\begin{equation}
\begin{aligned}
    \mathcal{Y}_{\mathcal{P}^{\mathrm{ent}}}
    (\phi,\theta,\Omega)
    &=
    C_B\left(
    \mathcal{P}^{\mathrm{ent}}
    (\ket{\phi,\gamma})
    \right)
    \\
    &=
    H\left(\{p_{ij}(\phi,\theta,\Omega)\}\right),
\end{aligned}
\label{eq:raw_yield_entangled}
\end{equation}
where $H$ denotes the Shannon entropy.

Equations~\eqref{eq:raw_yield_direct} and
\eqref{eq:raw_yield_entangled} establish the raw coherence yields of the two schemes. For coherent input states, while both constructions guarantee strictly positive yields, their relative performance varies based on the input state and the reference-frame misalignment. In the following subsections, we compare their achievable yields under two operational settings: phase optimization with knowledge of Bob's basis (Section~\ref{subsec:optimization}) and under unknown basis orientation (Section~\ref{subsec:average_yield}).

\begin{figure}[htbp]
    \centering
    \includegraphics[width=\linewidth]{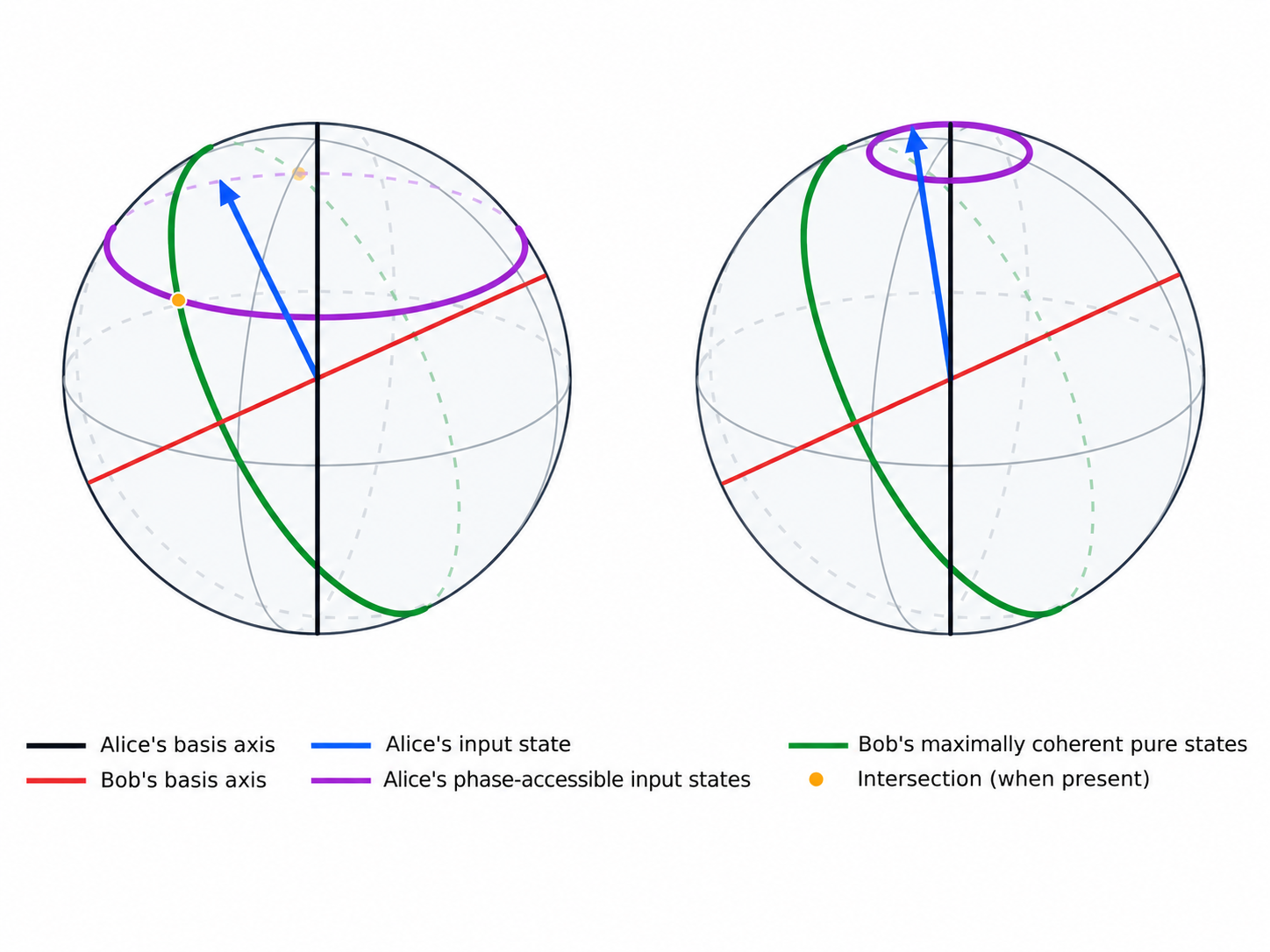}
    \caption{Bloch-sphere representation of the Direct Scheme optimization. The two panels depict scenarios with and without unit-yield solutions, respectively.}
    \label{fig:optimization_schematic}
\end{figure}

\begin{figure*}[t]
    \centering

    \begin{subfigure}[b]{0.46\textwidth}
        \centering
        \includegraphics[width=\linewidth]{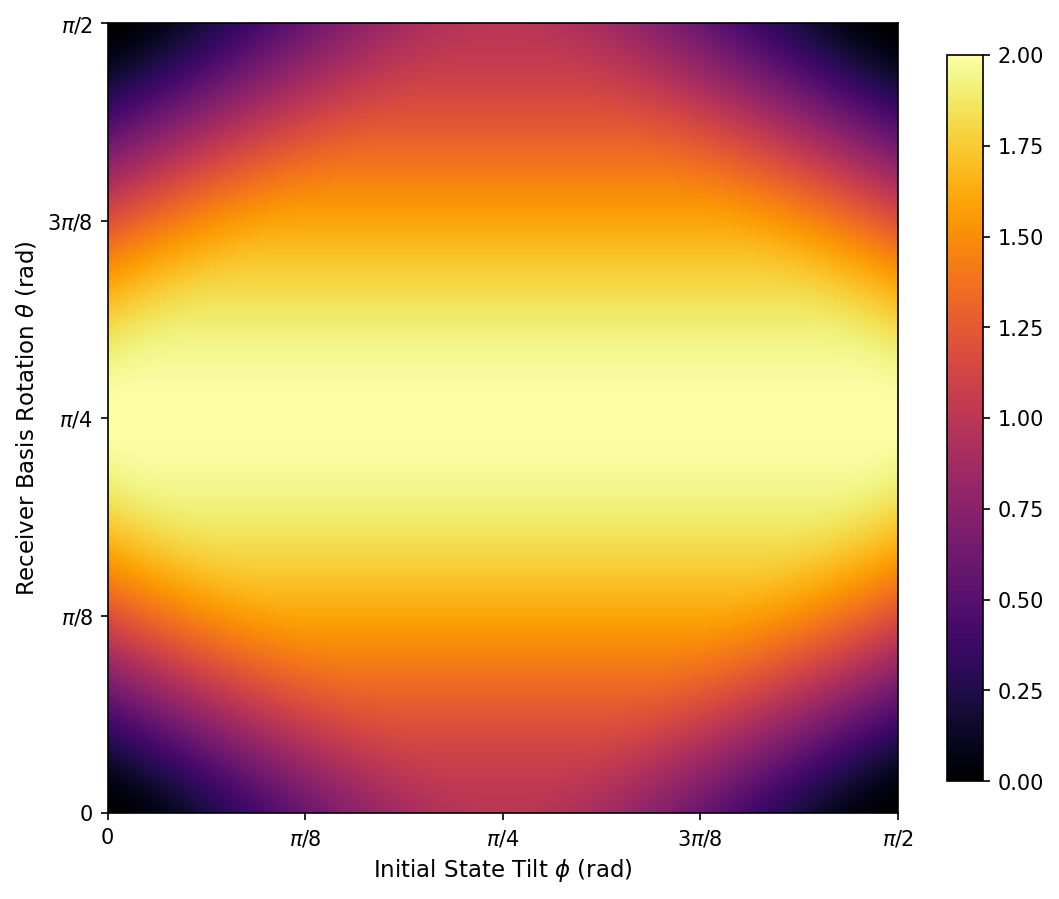}
        \caption{Direct Scheme}
        \label{fig:opt_yield_direct}
    \end{subfigure}
    \hfill
    \begin{subfigure}[b]{0.46\textwidth}
        \centering
        \includegraphics[width=\linewidth]{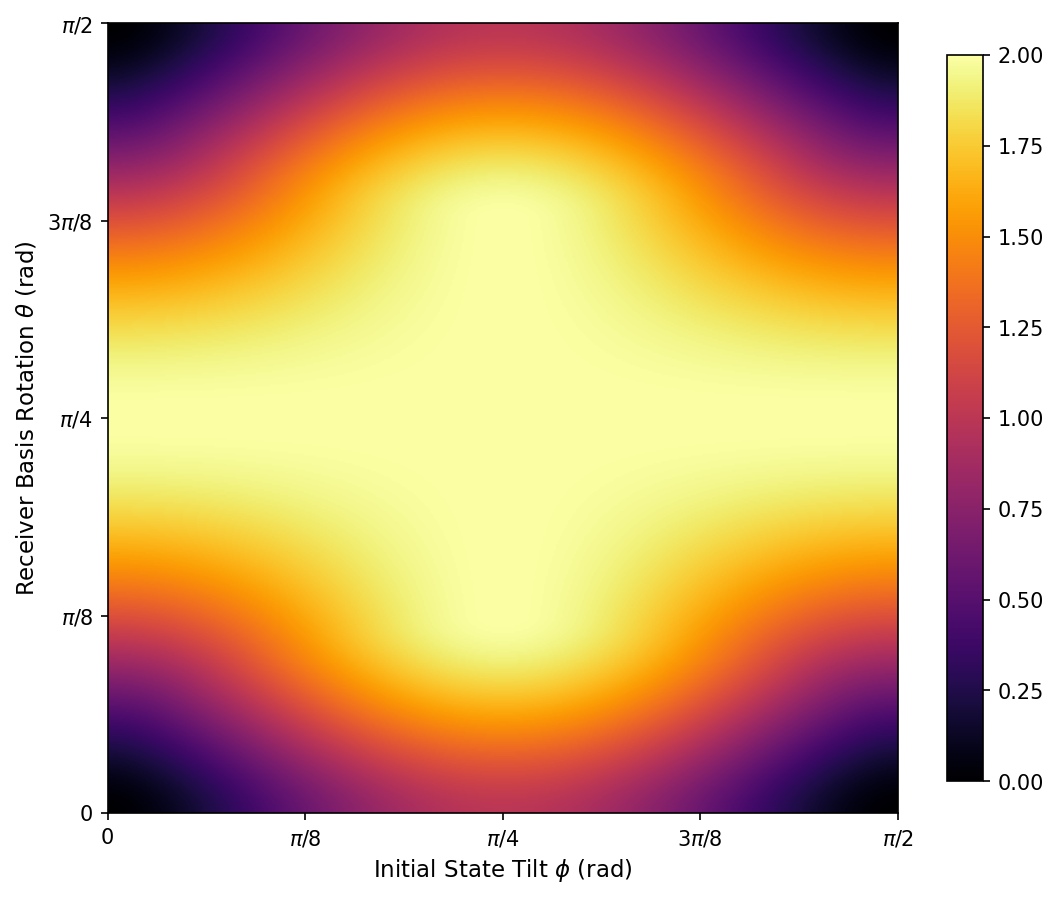}
        \caption{Entangled Scheme}
        \label{fig:opt_yield_entangled}
    \end{subfigure}

    \caption{Phase-optimized coherence yields for the
    (a) Direct Scheme and (b) Entangled Scheme,
    as functions of the initial state tilt $\phi$
    and receiver basis rotation $\theta$.}
    \label{fig:optimized_yields}
\end{figure*}

\subsection{Yield Optimization}
\label{subsec:optimization}
In this subsection, we evaluate the performance of both schemes under the operational assumption that Alice possesses full knowledge of Bob's reference basis parameters $(\theta,\delta)$. Since Alice may freely adjust the transmitted phase $\gamma'$, she can optimize the delivered yield over this parameter. In both schemes, the effective relative phase $\Omega$ depends linearly on $\gamma'$, so unrestricted optimization over $\gamma'$ is equivalent to optimization over $\Omega$. We therefore perform the following optimizations in terms of $\Omega$, with the corresponding optimal preparation phase $\gamma'$ determined by Alice from her knowledge of $\delta$.

For the Direct Scheme, phase optimization determines whether Alice can make the tilted qubit maximally coherent in Bob's reference basis. The corresponding optimal coherence-transfer yield is given below.

\begin{proposition}[Phase-Optimized Yield of the Direct Scheme]
\label{prop:direct_opt}
The phase-optimized coherence-transfer yield of the Direct Scheme is
\begin{widetext}
\begin{equation}
\label{eq:direct_optimal_yield}
\mathcal{Y}_{\mathcal{P}^{\mathrm{dir}}}^{*}(\phi,\theta)=
\begin{cases}
1+h(\cos^2\theta), & \text{if } \cos^2(\phi+\theta)\leq \frac12 \leq \cos^2(\phi-\theta), \\[2mm] \max\Big\{ h(\cos^2(\phi-\theta)), h(\cos^2(\phi+\theta)) \Big\} + h(\cos^2\theta), & \text{otherwise}.
\end{cases}
\end{equation}
\end{widetext}
\end{proposition}

\begin{proof}
See Appendix~\ref{app:direct_phase_optimization}.
\end{proof}

As illustrated in Fig.~\ref{fig:optimization_schematic}, the optimum depends on whether the phase-accessible states include a maximally coherent state in Bob's basis. Nevertheless, even when this maximum is inaccessible, the optimized Direct Scheme satisfies the following coherence guarantee.

\begin{figure}[htbp]
    \centering
    \includegraphics[width=\linewidth]{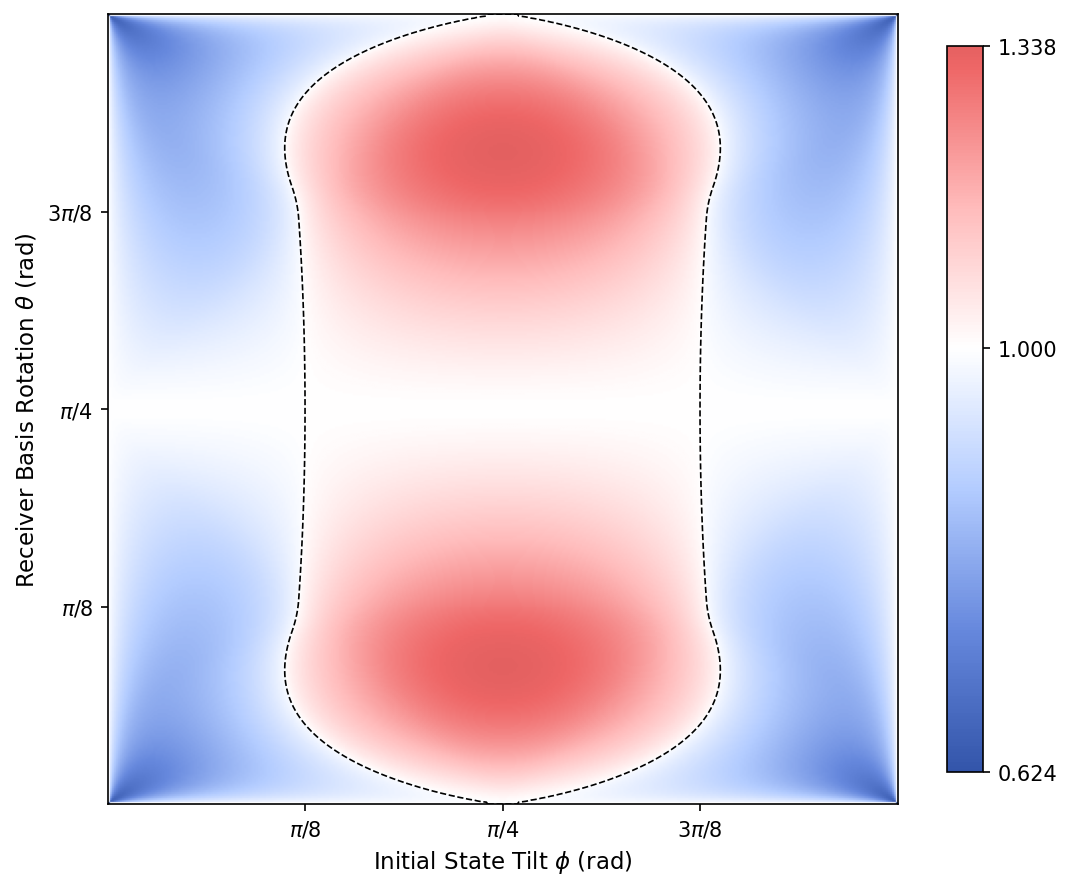}
    \caption{Ratio of phase-maximized coherence yields $\mathcal{Y}^*_{\mathcal{P}^{\mathrm{ent}}}(\phi, \theta) / \mathcal{Y}^*_{\mathcal{P}^{\mathrm{dir}}}(\phi, \theta)$ across state tilt $\phi$ and polar misalignment $\theta$. The dotted line traces the pairs for which the ratio is 1.}
    \label{fig:yield_ratio}
\end{figure}

\begin{observation}[Coherence Guarantee of the Direct Scheme]
\label{obs:direct_guarantee}
The phase-optimized Direct Scheme guarantees a coherence-transfer yield no smaller than Alice's initial coherence. Indeed, choosing $\gamma'=\delta+\pi/2$ eliminates the phase-dependent term in the tilted qubit probability, yielding
$p=(1+\cos(2\phi)\cos(2\theta))/2$. Since $|\cos(2\phi)\cos(2\theta)| \leq|\cos(2\phi)|$, the monotonicity of binary entropy gives
\begin{multline}
\label{dir_opt_lb}
\mathcal{Y}_{\mathcal{P}^{\mathrm{dir}}}^{*}(\phi,\theta)
\geq
\mathcal{Y}_{\mathcal{P}^{\mathrm{dir}}}
(\phi,\theta,\pi/4)\\
\geq h(\cos^2\phi)+h(\cos^2\theta)
\geq C_A(\ket{\phi,\gamma}).
\end{multline}
\end{observation}

We now turn to the Entangled Scheme, which guarantees a coherence-transfer yield at least as large as Alice's initial coherence even without phase optimization. When Alice additionally knows Bob's reference basis, optimizing the transmitted phase yields the following result.

\begin{proposition}[Phase-Optimized Yield of the Entangled Scheme]
\label{prop:entangled_opt}
The phase-optimized yield is
\begin{equation}
\label{eq:entangled_optimal_yield}
    \mathcal{Y}_{\mathcal{P}^{\mathrm{ent}}}^{*}
    (\phi,\theta)
    = H\big(\{p_{ij}(\Omega^*)\}\big),
\end{equation}
where the optimal phase $\Omega^*$ satisfies
\begin{equation}
\label{eq:entangled_optimal_phase}
\cos\Omega^* =
\begin{cases}
-\dfrac{\sin(2\phi)}{\tan^2(2\theta)},
& \tan^2(2\theta)>\sin(2\phi),\\[2mm]
-1, & \text{otherwise}.
\end{cases}
\end{equation}
\end{proposition}

\begin{proof}
See Appendix~\ref{app:entangled_phase_optimization}.
\end{proof}

\subsection{Average Yield Analysis}
\label{subsec:average_yield}

The preceding analysis considered the ideal scenario in which Alice possesses complete knowledge of Bob's reference basis parameters $(\theta,\delta)$, allowing her to optimize the transmitted phase. We now consider the complementary scenario in which Bob's reference basis is entirely unknown to Alice, as may occur in the absence of reference-frame synchronization or prior calibration. Similar considerations arise in reference-frame-independent Bell tests employing randomly chosen measurement settings~\cite{Cieśliński_2025}.

To evaluate the performance of both schemes under this uncertainty, we define the average coherence-transfer yield by uniformly averaging over Bob's reference basis orientations. For our parametrization of the Bloch sphere, the normalized Haar-induced measure is
\begin{equation}
    d\mu(\theta,\delta)
    =\frac{\sin(2\theta)}{2\pi}
    \,d\theta\,d\delta,
\end{equation}
where $\theta\in[0,\pi/2]$ and $\delta\in[0,2\pi)$ \cite{Harr_Measures}.

\begin{definition}[Average Coherence-Transfer Yield]
\label{def:average_yield}
The Haar-averaged coherence-transfer yield of a protocol
$\mathcal{P}$ is
\begin{equation}
\label{eq:avg_yield_def}
\begin{aligned}
    \left\langle\mathcal{Y}_{\mathcal{P}}
    \right\rangle_{\theta,\delta}(\phi)
    =\frac{1}{2\pi}
    \int_0^{2\pi}d\delta
    \int_0^{\pi/2}d\theta\,
    \sin(2\theta)\,
    \mathcal{Y}_{\mathcal{P}}
    (\phi,\theta,\Omega).
\end{aligned}
\end{equation}
\end{definition}

Although the effective phase $\Omega$ depends on both $\gamma'$ and $\delta$ in both schemes, the periodicity of the yield functions ensures that uniform averaging over $\delta$ is equivalent to averaging over a complete period of $\Omega$ in either scheme. Consequently, the expected yield is independent of Alice's chosen phase $\gamma'$. This contrasts with the preceding optimization, where Alice exploits her knowledge of $\delta$ to select the phase that maximizes the delivered coherence.

For the Direct Scheme, the average yield admits an exact analytical expression.

\begin{proposition}[Average Yield of the Direct Scheme]
\label{prop:direct_average}
The Haar-averaged coherence-transfer yield of the Direct Scheme is independent of Alice's pure input state and satisfies
\begin{equation}
\label{eq:direct_average_yield}
    \left\langle
    \mathcal{Y}_{\mathcal{P}^{\mathrm{dir}}}
    \right\rangle_{\theta,\delta}(\phi)
    =\frac{1}{\ln 2}.
\end{equation}
\end{proposition}

\begin{proof}
Let $\ket{\psi}$ be an arbitrary fixed pure qubit and $\mathcal{B}_U=\{U\ket{0},U\ket{1}\}$ a Haar-random reference basis. The probability $p_{\psi}=|\bra{0}U^\dagger\ket{\psi}|^2$ has a distribution independent of $\ket{\psi}$: writing $\ket{\psi}=V\ket{0}$ for some unitary $V$, this follows from the invariance of the Haar measure under $U\mapsto V^\dagger U$. By Haar invariance, the distribution of \(p_\psi\) is therefore identical to that of \(p_0 = \cos^2\theta\) for the reference state \(\ket{0}\). Since \(dp_0=-\sin(2\theta)d\theta\) under this parametrization, the Haar-induced distribution of \(p_\psi\) is uniform on \([0,1]\) \cite{random_states}.
\begin{equation}
\begin{aligned}
    \left\langle
    C_{\mathcal{B}_U}(\ket{\psi})
    \right\rangle_U
    &=\int_0^1 h(p)\,dp\\
    &=\frac{1}{2\ln 2}.
\end{aligned}
\end{equation}
Since the Direct Scheme transmits two pure product qubits,
additivity of the relative entropy of coherence gives
Eq.~\eqref{eq:direct_average_yield}.
\end{proof}

\begin{figure}[htbp]
    \centering
    \includegraphics[width=\linewidth]{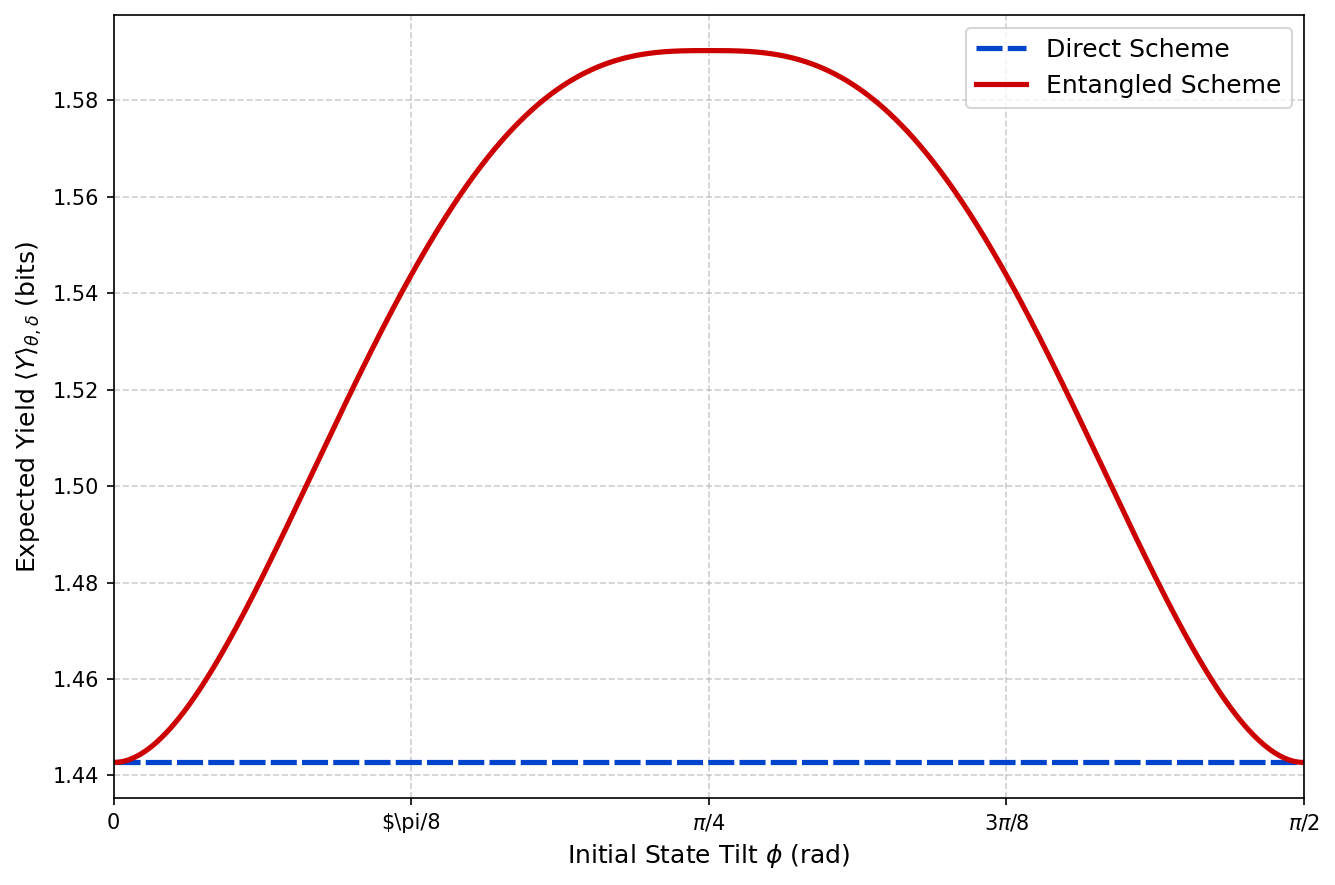}
    \caption{Haar-averaged coherence-transfer yields of
    the Direct Scheme (blue) and Entangled Scheme (red)
    as functions of Alice's input-state tilt $\phi$.}
    \label{fig:average_yield}
\end{figure}

For the Entangled Scheme, the average yield is evaluated numerically by integrating the raw yield according to Eq.~\eqref{eq:avg_yield_def}. The resulting average yields of both schemes are compared in Fig.~\ref{fig:average_yield}.

The results reveal a strict average-case advantage for the Entangled Scheme whenever Alice's input state is coherent. While the Direct Scheme maintains a constant average yield of $1/\ln 2\approx1.4427$ bits per channel use, the Entangled Scheme exhibits a nontrivial dependence on $\phi$, reaching a maximum average yield of approximately $1.5903$ bits per channel use.

The two schemes have equal average yields at the incoherent input configurations $\phi=0$ and $\phi=\pi/2$. At these points, the CNOT operation does not generate entanglement, and both schemes transmit product states of local incoherent basis vectors. Their average yields therefore coincide, as illustrated in Fig.~\ref{fig:average_yield}.

This behavior contrasts with the phase-optimized scenario, where both schemes exhibit regions of relative advantage depending on the input state and Bob's basis orientation. Under complete basis uncertainty, the numerical results instead favor the Entangled Scheme for every coherent input state.

\section{Asymptotic Coherence Transfer}
\label{sec:asymptotic}

We now investigate coherence transfer in the asymptotic regime, where Alice has access to an arbitrarily large number of independent copies of her initial state and quantum channel uses. Unlike the single-shot setting, the asymptotic reversibility of coherence under maximally incoherent operations (MIO) allows us to characterize the available resources entirely in terms of the initial relative entropy of coherence, substantially simplifying the optimization problem.

It is known that the asymptotic distillable coherence and coherence cost under MIO coincide and are both quantified by the relative entropy of coherence \cite{DIO_reversibility, one-shot-dilution}:
\begin{equation}
    C_D^{\mathrm{MIO}}(\rho)
    =
    C_C^{\mathrm{MIO}}(\rho)
    =
    C_A(\rho).
\end{equation}
Consequently, for any two states $\rho$ and $\sigma$ with nonzero coherence, asymptotic state interconversion under MIO is possible at the optimal rate
\begin{equation}
    R(\rho\rightarrow\sigma)
    =
    \frac{C_A(\rho)}{C_A(\sigma)}.
\end{equation}
These transformations are understood to be asymptotically faithful, with approximation errors vanishing in the many-copy limit.

Consider a sequence of communication scenarios in which the number of input copies $m$ and channel uses $n$ both diverge, while their ratio approaches a fixed value
\begin{equation}
    r=\lim_{n\rightarrow\infty}\frac{m}{n}.
\end{equation}
Alice therefore possesses an asymptotic coherence budget of $rC_A(\rho)$ bits per channel use. By reversibility, any two input states with identical relative entropy of coherence are asymptotically interconvertible at unit rate under MIO.  Consequently, the asymptotic optimal coherence-transfer yield depends on the initial state only through its relative entropy of coherence, denoted by $C_A(\rho)\equiv C$.

\begin{definition}[Asymptotic Optimal Coherence-Transfer Yield]
\label{def:asymptotic_yield}
For a fixed input-copy rate $r$, the asymptotic optimal
coherence-transfer yield is
\begin{equation}
    \widehat{\mathcal{Y}}_Q^\infty(C;r)
    :=
    \limsup_{n\to\infty}
    \widehat{\mathcal{Y}}_Q^{n,m(n)}(\rho),
\end{equation}
where $m(n)/n\to r$ and $C_A(\rho)\equiv C$. For the natural choice $m=n$, we often suppress the copy-rate parameter and write $\widehat{\mathcal{Y}}_Q^\infty(C)$.
\end{definition}

\begin{observation}[Capacity Saturation]
\label{obs:capacity_saturation}
The asymptotic optimal coherence-transfer yield saturates the communication bound whenever Alice's available coherence per channel use satisfies $rC\geq Q$. Specifically, for $C>0$,
\begin{equation}
    \widehat{\mathcal{Y}}_Q^\infty(C;r) = \widehat{\mathcal{Y}}_Q^\infty(C;Q/C) = Q,
    \qquad r\geq Q/C,
\end{equation}
provided the channel's full dimensional capacity can be utilized asymptotically. Indeed, asymptotic reversibility allows Alice to prepare maximally coherent output states whose coherence cost does not exceed the available budget $rC$. The channel-capacity bound then establishes equality.
\end{observation}

Thus, the nontrivial optimization is restricted to the regime $rC<Q$. In the following, we specialize to the case $r=1$ and $Q=1$, where Alice and Bob are equipped with qubit coherence theories and Alice initially possesses $n$ copies of a qubit state $\rho$, with $C_A(\rho)=C\leq 1$. We investigate achievable coherence-transfer yields in this setting by exploiting the freedom afforded by asymptotic MIO reversibility.

Consider the following restricted family of asymptotic protocols in which Alice converts her initial state into $n$ identical copies of a single-qubit target state $\sigma$, which are subsequently transmitted to Bob.

\begin{definition}[Asymptotic IID Coherence-Transfer Yields]
\label{def:asymptotic_iid_yields}
We define the achievable coherence-transfer yields for arbitrary single-qubit targets and their pure-state restriction, respectively, as
\begin{equation}
\begin{aligned}
G_\theta(C)
&:=\max_{\substack{\sigma\in\mathcal S(\mathbb C^2)\\
C_A(\sigma)\leq C}}C_B(\sigma),\\
F_\theta(C)
&:=\max_{\substack{\ket{\psi}\in\mathbb C^2\\
C_A(\ket{\psi}\bra{\psi})\leq C}}C_B(\ket{\psi}).
\end{aligned}
\label{eq:asymptotic_iid_bounds}
\end{equation}
\end{definition}

The achievability of these quantities follows from asymptotic MIO reversibility, which permits Alice to prepare the corresponding target states at unit rate whenever their coherence cost does not exceed $C$. By construction, $F_\theta(C)\leq G_\theta(C)\leq\widehat{\mathcal Y}_1^\infty(C)$.

We first evaluate $F_\theta(C)$ analytically. Parameterizing the pure target as $\ket{\phi,\gamma'}=\cos\phi\ket{0} +e^{i\gamma'}\sin\phi\ket{1}$, its coherence cost is $h(\cos^2\phi)$. The delivered coherence is precisely the single-qubit contribution to the Direct Scheme derived in Sec.~\ref{sec:scenario_raw_yields}. Unlike the single-shot setting, Alice can now optimize over both the coherence of the target state and its phase.

\begin{proposition}[Optimal Pure-State Yield]
\label{prop:asymptotic_pure}
The optimal pure-state yield is
\begin{equation}
    F_\theta(C)
    =
    \begin{cases}
        1, & C\geq c_*(\theta),\\[4pt]
        \mathcal{Y}^*_{\mathcal{P}^{\mathrm{dir}}}(\phi_C,\theta) - h(\cos^2 \theta),
        & C<c_*(\theta).
    \end{cases}
\label{eq:asymptotic_combined_bound}
\end{equation}
where $\phi_C\in[0,\pi/4]$ satisfies $h(\cos^2\phi_C)=C$, and
\begin{equation}
\begin{aligned}
c_*(\theta)
&:=\min_{\ket{\psi}:\,C_B(\ket{\psi})=1}C_A(\ket{\psi})\\
&=h\left(\frac{1+|\sin(2\theta)|}{2}\right).
\end{aligned}
\label{eq:asymptotic_threshold}
\end{equation}
\end{proposition}

\begin{proof}
The optimization extends the phase-optimized Direct Scheme by allowing Alice to vary the target state's coherence in addition to its phase. The threshold $c_*(\theta)$ is the minimum Alice-coherence cost required for the phase-accessible states to include a maximally coherent state in Bob's basis, corresponding to the unit-yield solution of Proposition~\ref{prop:direct_opt}. Above this threshold, asymptotic MIO reversibility allows Alice to achieve unit yield. Below it, we show in Appendix~\ref{app:asymptotic_optimization} that the optimal pure-state target exhausts Alice's coherence budget. The result then follows from the phase-optimized Direct Scheme yield, evaluated at $\phi=\phi_C$.
\end{proof}

Since $F_\theta(C)=1$ saturates the channel-capacity bound, we also obtain
\begin{equation}
\label{eq:asymptotic_unit_yield}
\widehat{\mathcal Y}_1^\infty(C)=1,
\qquad C\geq c_*(\theta).
\end{equation}

\begin{figure}[h]
    \centering
    \includegraphics[width=\linewidth]{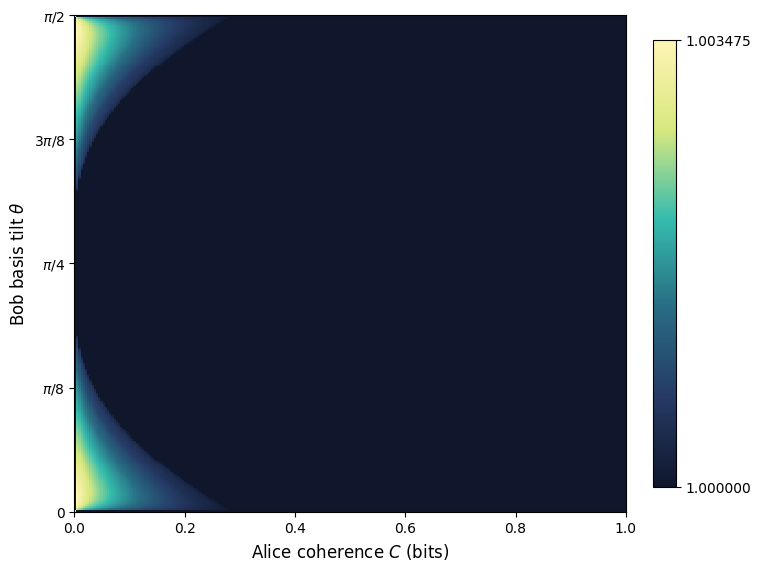}
    \caption{Ratio of the optimized mixed-state and pure-state asymptotic coherence yields, $G_\theta(C)/F_\theta(C)$, as a function of Alice's initial coherence $C$ and the relative basis tilt $\theta$.}
    \label{fig:asymptotic_ratio}
\end{figure}

We then investigate the more general single-qubit optimization $G_\theta(C)$. Representing the target states by Bloch vectors, optimization over their azimuthal orientation can be performed analytically as shown in Appendix~\ref{app:asymptotic_optimization}. We then perform a numerical optimization over the remaining parameters. As shown in Fig.~\ref{fig:asymptotic_ratio}, mixed targets can slightly outperform the pure-state construction in a restricted region of low input coherence and small relative basis misalignment. However, the improvement remains modest: the maximum observed difference $G_\theta(C)-F_\theta(C)$ over the sampled parameter grid is approximately $4.2\times10^{-4}$ bits.

By an argument analogous to that of Eq.~\eqref{dir_opt_lb}, the phase-optimized pure-state protocol guarantees a coherence-transfer yield at least as large as Alice's initial coherence. Combining this with the preceding results, we obtain
\begin{equation}
    C
    \leq F_\theta(C)
    \leq G_\theta(C)
    \leq \widehat{\mathcal{Y}}_1^\infty(C)
    \leq 1.
\end{equation}
While the pure-state protocol is globally optimal in the capacity-saturating regime and numerically close to the general IID single-qubit optimum across the investigated parameter range, the possibility of improvements through nonidentical or correlated asymptotic output states remains open.

\section{Conclusion}
\label{sec:conclusion}

We have investigated quantum coherence transfer between two parties with misaligned local reference bases under a constraint on quantum communication capacity. Within this framework, we analyzed two single-shot strategies based on direct transmission and coherence-to-entanglement conversion, deriving their optimized coherence yields. We also examined their performance when Bob's reference basis is unknown, using Haar averaging to characterize the expected coherence transfer over possible basis configurations.

In the asymptotic regime, we exploited the reversibility of coherence under MIO to formulate the transfer problem in terms of Alice's initial coherence. We derived an analytical achievable yield based on pure-state conversions and identified a regime in which this construction saturates the communication capacity, thereby establishing optimality. Numerical optimization over general single-qubit target states further revealed small improvements over the pure-state construction.

Several directions remain open for further investigation. A natural extension is to replace the ideal quantum communication channel with a general noisy channel and investigate how resource-theoretic restrictions on encoding and decoding affect the achievable coherence-transfer rates \cite{channel_coherence}. Another question concerns the optimality of independent single-qubit preparations in the asymptotic regime, where more general correlated output strategies may offer an advantage. Finally, extending the analysis to scenarios in which Alice has only partial information about Bob's reference basis could clarify the role of reference-frame knowledge in coherence transfer.

\begin{acknowledgments}
KH acknowledges National Science Centre grant Opus 25 no. UMO-2023/49/B/ST2/02468. The authors acknowledge the use of ChatGPT (OpenAI), powered by GPT-6, for assistance with scientific writing and editing, numerical code development, and review of mathematical arguments. All AI-assisted material was reviewed and verified by the authors, who retain full responsibility for the scientific results and the content of this manuscript.
\end{acknowledgments}

\bibliography{apssamp.bib}

\begin{thebibliography}{28}%
\makeatletter
\providecommand \@ifxundefined [1]{%
 \@ifx{#1\undefined}
}%
\providecommand \@ifnum [1]{%
 \ifnum #1\expandafter \@firstoftwo
 \else \expandafter \@secondoftwo
 \fi
}%
\providecommand \@ifx [1]{%
 \ifx #1\expandafter \@firstoftwo
 \else \expandafter \@secondoftwo
 \fi
}%
\providecommand \natexlab [1]{#1}%
\providecommand \enquote  [1]{``#1''}%
\providecommand \bibnamefont  [1]{#1}%
\providecommand \bibfnamefont [1]{#1}%
\providecommand \citenamefont [1]{#1}%
\providecommand \href@noop [0]{\@secondoftwo}%
\providecommand \href [0]{\begingroup \@sanitize@url \@href}%
\providecommand \@href[1]{\@@startlink{#1}\@@href}%
\providecommand \@@href[1]{\endgroup#1\@@endlink}%
\providecommand \@sanitize@url [0]{\catcode `\\12\catcode `\$12\catcode `\&12\catcode `\#12\catcode `\^12\catcode `\_12\catcode `\%12\relax}%
\providecommand \@@startlink[1]{}%
\providecommand \@@endlink[0]{}%
\providecommand \url  [0]{\begingroup\@sanitize@url \@url }%
\providecommand \@url [1]{\endgroup\@href {#1}{\urlprefix }}%
\providecommand \urlprefix  [0]{URL }%
\providecommand \Eprint [0]{\href }%
\providecommand \doibase [0]{https://doi.org/}%
\providecommand \selectlanguage [0]{\@gobble}%
\providecommand \bibinfo  [0]{\@secondoftwo}%
\providecommand \bibfield  [0]{\@secondoftwo}%
\providecommand \translation [1]{[#1]}%
\providecommand \BibitemOpen [0]{}%
\providecommand \bibitemStop [0]{}%
\providecommand \bibitemNoStop [0]{.\EOS\space}%
\providecommand \EOS [0]{\spacefactor3000\relax}%
\providecommand \BibitemShut  [1]{\csname bibitem#1\endcsname}%
\let\auto@bib@innerbib\@empty
\bibitem [{\citenamefont {Horodecki}\ \emph {et~al.}(2009)\citenamefont {Horodecki}, \citenamefont {Horodecki}, \citenamefont {Horodecki},\ and\ \citenamefont {Horodecki}}]{Horodecki2009_Entanglement}%
  \BibitemOpen
  \bibfield  {author} {\bibinfo {author} {\bibfnamefont {R.}~\bibnamefont {Horodecki}}, \bibinfo {author} {\bibfnamefont {P.}~\bibnamefont {Horodecki}}, \bibinfo {author} {\bibfnamefont {M.}~\bibnamefont {Horodecki}},\ and\ \bibinfo {author} {\bibfnamefont {K.}~\bibnamefont {Horodecki}},\ }\bibfield  {title} {\bibinfo {title} {Quantum entanglement},\ }\href {https://doi.org/10.1103/RevModPhys.81.865} {\bibfield  {journal} {\bibinfo  {journal} {Rev. Mod. Phys.}\ }\textbf {\bibinfo {volume} {81}},\ \bibinfo {pages} {865} (\bibinfo {year} {2009})}\BibitemShut {NoStop}%
\bibitem [{\citenamefont {Glauber}(1963)}]{Glauber1963}%
  \BibitemOpen
  \bibfield  {author} {\bibinfo {author} {\bibfnamefont {R.~J.}\ \bibnamefont {Glauber}},\ }\bibfield  {title} {\bibinfo {title} {Coherent and incoherent states of the radiation field},\ }\href {https://doi.org/10.1103/PhysRev.131.2766} {\bibfield  {journal} {\bibinfo  {journal} {Phys. Rev.}\ }\textbf {\bibinfo {volume} {131}},\ \bibinfo {pages} {2766} (\bibinfo {year} {1963})}\BibitemShut {NoStop}%
\bibitem [{\citenamefont {Mandel}\ and\ \citenamefont {Wolf}(1995)}]{MandelWolf1995}%
  \BibitemOpen
  \bibfield  {author} {\bibinfo {author} {\bibfnamefont {L.}~\bibnamefont {Mandel}}\ and\ \bibinfo {author} {\bibfnamefont {E.}~\bibnamefont {Wolf}},\ }\href {https://www.cambridge.org/core/books/optical-coherence-and-quantum-optics/F8CB94C70FA64CD3FB60890CA2048168} {\emph {\bibinfo {title} {{Optical Coherence and Quantum Optics}}}}\ (\bibinfo  {publisher} {Cambridge University Press},\ \bibinfo {address} {Cambridge},\ \bibinfo {year} {1995})\BibitemShut {NoStop}%
\bibitem [{\citenamefont {Baumgratz}\ \emph {et~al.}(2014)\citenamefont {Baumgratz}, \citenamefont {Cramer},\ and\ \citenamefont {Plenio}}]{Baumgratz2014_Coherence}%
  \BibitemOpen
  \bibfield  {author} {\bibinfo {author} {\bibfnamefont {T.}~\bibnamefont {Baumgratz}}, \bibinfo {author} {\bibfnamefont {M.}~\bibnamefont {Cramer}},\ and\ \bibinfo {author} {\bibfnamefont {M.~B.}\ \bibnamefont {Plenio}},\ }\bibfield  {title} {\bibinfo {title} {Quantifying coherence},\ }\href {https://doi.org/10.1103/PhysRevLett.113.140401} {\bibfield  {journal} {\bibinfo  {journal} {Phys. Rev. Lett.}\ }\textbf {\bibinfo {volume} {113}},\ \bibinfo {pages} {140401} (\bibinfo {year} {2014})}\BibitemShut {NoStop}%
\bibitem [{\citenamefont {Streltsov}\ \emph {et~al.}(2017{\natexlab{a}})\citenamefont {Streltsov}, \citenamefont {Adesso},\ and\ \citenamefont {Plenio}}]{Streltsov2017_CoherenceReview}%
  \BibitemOpen
  \bibfield  {author} {\bibinfo {author} {\bibfnamefont {A.}~\bibnamefont {Streltsov}}, \bibinfo {author} {\bibfnamefont {G.}~\bibnamefont {Adesso}},\ and\ \bibinfo {author} {\bibfnamefont {M.~B.}\ \bibnamefont {Plenio}},\ }\bibfield  {title} {\bibinfo {title} {Colloquium: Quantum coherence as a resource},\ }\href {https://doi.org/10.1103/RevModPhys.89.041003} {\bibfield  {journal} {\bibinfo  {journal} {Rev. Mod. Phys.}\ }\textbf {\bibinfo {volume} {89}},\ \bibinfo {pages} {041003} (\bibinfo {year} {2017}{\natexlab{a}})}\BibitemShut {NoStop}%
\bibitem [{\citenamefont {Levi}\ and\ \citenamefont {Mintert}(2014)}]{Levi_2014}%
  \BibitemOpen
  \bibfield  {author} {\bibinfo {author} {\bibfnamefont {F.}~\bibnamefont {Levi}}\ and\ \bibinfo {author} {\bibfnamefont {F.}~\bibnamefont {Mintert}},\ }\bibfield  {title} {\bibinfo {title} {A quantitative theory of coherent delocalization},\ }\href {https://doi.org/10.1088/1367-2630/16/3/033007} {\bibfield  {journal} {\bibinfo  {journal} {New Journal of Physics}\ }\textbf {\bibinfo {volume} {16}},\ \bibinfo {pages} {033007} (\bibinfo {year} {2014})}\BibitemShut {NoStop}%
\bibitem [{\citenamefont {Aberg}(2006)}]{aberg2006}%
  \BibitemOpen
  \bibfield  {author} {\bibinfo {author} {\bibfnamefont {J.}~\bibnamefont {Aberg}},\ }\href {https://arxiv.org/abs/quant-ph/0612146} {\bibinfo {title} {Quantifying superposition}} (\bibinfo {year} {2006}),\ \Eprint {https://arxiv.org/abs/quant-ph/0612146} {arXiv:quant-ph/0612146 [quant-ph]} \BibitemShut {NoStop}%
\bibitem [{\citenamefont {Marvian}\ and\ \citenamefont {Spekkens}(2016)}]{speakable_unspeakable}%
  \BibitemOpen
  \bibfield  {author} {\bibinfo {author} {\bibfnamefont {I.}~\bibnamefont {Marvian}}\ and\ \bibinfo {author} {\bibfnamefont {R.~W.}\ \bibnamefont {Spekkens}},\ }\bibfield  {title} {\bibinfo {title} {How to quantify coherence: Distinguishing speakable and unspeakable notions},\ }\href {https://doi.org/10.1103/PhysRevA.94.052324} {\bibfield  {journal} {\bibinfo  {journal} {Phys. Rev. A}\ }\textbf {\bibinfo {volume} {94}},\ \bibinfo {pages} {052324} (\bibinfo {year} {2016})}\BibitemShut {NoStop}%
\bibitem [{\citenamefont {Gour}\ and\ \citenamefont {Spekkens}(2008)}]{Gour_2008}%
  \BibitemOpen
  \bibfield  {author} {\bibinfo {author} {\bibfnamefont {G.}~\bibnamefont {Gour}}\ and\ \bibinfo {author} {\bibfnamefont {R.~W.}\ \bibnamefont {Spekkens}},\ }\bibfield  {title} {\bibinfo {title} {The resource theory of quantum reference frames: manipulations and monotones},\ }\href {https://doi.org/10.1088/1367-2630/10/3/033023} {\bibfield  {journal} {\bibinfo  {journal} {New Journal of Physics}\ }\textbf {\bibinfo {volume} {10}},\ \bibinfo {pages} {033023} (\bibinfo {year} {2008})}\BibitemShut {NoStop}%
\bibitem [{\citenamefont {Bartlett}\ \emph {et~al.}(2003)\citenamefont {Bartlett}, \citenamefont {Rudolph},\ and\ \citenamefont {Spekkens}}]{Bartlett2007_ReferenceFrames}%
  \BibitemOpen
  \bibfield  {author} {\bibinfo {author} {\bibfnamefont {S.~D.}\ \bibnamefont {Bartlett}}, \bibinfo {author} {\bibfnamefont {T.}~\bibnamefont {Rudolph}},\ and\ \bibinfo {author} {\bibfnamefont {R.~W.}\ \bibnamefont {Spekkens}},\ }\bibfield  {title} {\bibinfo {title} {Classical and quantum communication without a shared reference frame},\ }\href {https://doi.org/10.1103/PhysRevLett.91.027901} {\bibfield  {journal} {\bibinfo  {journal} {Phys. Rev. Lett.}\ }\textbf {\bibinfo {volume} {91}},\ \bibinfo {pages} {027901} (\bibinfo {year} {2003})}\BibitemShut {NoStop}%
\bibitem [{\citenamefont {Laing}\ \emph {et~al.}(2010)\citenamefont {Laing}, \citenamefont {Scarani}, \citenamefont {Rarity},\ and\ \citenamefont {O'Brien}}]{Laing2010_RFI}%
  \BibitemOpen
  \bibfield  {author} {\bibinfo {author} {\bibfnamefont {A.}~\bibnamefont {Laing}}, \bibinfo {author} {\bibfnamefont {V.}~\bibnamefont {Scarani}}, \bibinfo {author} {\bibfnamefont {J.~G.}\ \bibnamefont {Rarity}},\ and\ \bibinfo {author} {\bibfnamefont {J.~L.}\ \bibnamefont {O'Brien}},\ }\bibfield  {title} {\bibinfo {title} {Reference-frame-independent quantum key distribution},\ }\href {https://doi.org/10.1103/PhysRevA.82.012304} {\bibfield  {journal} {\bibinfo  {journal} {Phys. Rev. A}\ }\textbf {\bibinfo {volume} {82}},\ \bibinfo {pages} {012304} (\bibinfo {year} {2010})}\BibitemShut {NoStop}%
\bibitem [{\citenamefont {D’Ambrosio}\ \emph {et~al.}(2012)\citenamefont {D’Ambrosio}, \citenamefont {Nagali}, \citenamefont {Walborn}, \citenamefont {Aolita}, \citenamefont {Slussarenko}, \citenamefont {Marrucci},\ and\ \citenamefont {Sciarrino}}]{DAmbrosio2012_FrameDrift}%
  \BibitemOpen
  \bibfield  {author} {\bibinfo {author} {\bibfnamefont {V.}~\bibnamefont {D’Ambrosio}}, \bibinfo {author} {\bibfnamefont {E.}~\bibnamefont {Nagali}}, \bibinfo {author} {\bibfnamefont {S.~P.}\ \bibnamefont {Walborn}}, \bibinfo {author} {\bibfnamefont {L.}~\bibnamefont {Aolita}}, \bibinfo {author} {\bibfnamefont {S.}~\bibnamefont {Slussarenko}}, \bibinfo {author} {\bibfnamefont {L.}~\bibnamefont {Marrucci}},\ and\ \bibinfo {author} {\bibfnamefont {F.}~\bibnamefont {Sciarrino}},\ }\bibfield  {title} {\bibinfo {title} {Complete experimental toolbox for alignment-free quantum communication},\ }\bibfield  {journal} {\bibinfo  {journal} {Nature Communications}\ }\textbf {\bibinfo {volume} {3}},\ \href {https://doi.org/10.1038/ncomms1951} {10.1038/ncomms1951} (\bibinfo {year} {2012})\BibitemShut {NoStop}%
\bibitem [{\citenamefont {Chen}\ \emph {et~al.}(2006)\citenamefont {Chen}, \citenamefont {Zhang}, \citenamefont {Boileau}, \citenamefont {Jin}, \citenamefont {Yang}, \citenamefont {Zhang}, \citenamefont {Yang}, \citenamefont {Laflamme},\ and\ \citenamefont {Pan}}]{experiment_qc}%
  \BibitemOpen
  \bibfield  {author} {\bibinfo {author} {\bibfnamefont {T.-Y.}\ \bibnamefont {Chen}}, \bibinfo {author} {\bibfnamefont {J.}~\bibnamefont {Zhang}}, \bibinfo {author} {\bibfnamefont {J.-C.}\ \bibnamefont {Boileau}}, \bibinfo {author} {\bibfnamefont {X.-M.}\ \bibnamefont {Jin}}, \bibinfo {author} {\bibfnamefont {B.}~\bibnamefont {Yang}}, \bibinfo {author} {\bibfnamefont {Q.}~\bibnamefont {Zhang}}, \bibinfo {author} {\bibfnamefont {T.}~\bibnamefont {Yang}}, \bibinfo {author} {\bibfnamefont {R.}~\bibnamefont {Laflamme}},\ and\ \bibinfo {author} {\bibfnamefont {J.-W.}\ \bibnamefont {Pan}},\ }\bibfield  {title} {\bibinfo {title} {Experimental quantum communication without a shared reference frame},\ }\href {https://doi.org/10.1103/PhysRevLett.96.150504} {\bibfield  {journal} {\bibinfo  {journal} {Phys. Rev. Lett.}\ }\textbf {\bibinfo {volume} {96}},\ \bibinfo {pages} {150504} (\bibinfo {year} {2006})}\BibitemShut {NoStop}%
\bibitem [{\citenamefont {Winter}\ and\ \citenamefont {Yang}(2016)}]{RT_Coherence}%
  \BibitemOpen
  \bibfield  {author} {\bibinfo {author} {\bibfnamefont {A.}~\bibnamefont {Winter}}\ and\ \bibinfo {author} {\bibfnamefont {D.}~\bibnamefont {Yang}},\ }\bibfield  {title} {\bibinfo {title} {Operational resource theory of coherence},\ }\href {https://doi.org/10.1103/PhysRevLett.116.120404} {\bibfield  {journal} {\bibinfo  {journal} {Phys. Rev. Lett.}\ }\textbf {\bibinfo {volume} {116}},\ \bibinfo {pages} {120404} (\bibinfo {year} {2016})}\BibitemShut {NoStop}%
\bibitem [{\citenamefont {Chitambar}\ and\ \citenamefont {Gour}(2016{\natexlab{a}})}]{ChitambarGour2016}%
  \BibitemOpen
  \bibfield  {author} {\bibinfo {author} {\bibfnamefont {E.}~\bibnamefont {Chitambar}}\ and\ \bibinfo {author} {\bibfnamefont {G.}~\bibnamefont {Gour}},\ }\bibfield  {title} {\bibinfo {title} {Critical examination of incoherent operations and a physically consistent resource theory of quantum coherence},\ }\href {https://doi.org/10.1103/PhysRevLett.117.030401} {\bibfield  {journal} {\bibinfo  {journal} {Phys. Rev. Lett.}\ }\textbf {\bibinfo {volume} {117}},\ \bibinfo {pages} {030401} (\bibinfo {year} {2016}{\natexlab{a}})}\BibitemShut {NoStop}%
\bibitem [{\citenamefont {Chitambar}\ and\ \citenamefont {Gour}(2016{\natexlab{b}})}]{ChitambarGour2016_Comparison}%
  \BibitemOpen
  \bibfield  {author} {\bibinfo {author} {\bibfnamefont {E.}~\bibnamefont {Chitambar}}\ and\ \bibinfo {author} {\bibfnamefont {G.}~\bibnamefont {Gour}},\ }\bibfield  {title} {\bibinfo {title} {Comparison of incoherent operations and measures of coherence},\ }\href {https://doi.org/10.1103/PhysRevA.94.052336} {\bibfield  {journal} {\bibinfo  {journal} {Phys. Rev. A}\ }\textbf {\bibinfo {volume} {94}},\ \bibinfo {pages} {052336} (\bibinfo {year} {2016}{\natexlab{b}})}\BibitemShut {NoStop}%
\bibitem [{\citenamefont {Streltsov}\ \emph {et~al.}(2017{\natexlab{b}})\citenamefont {Streltsov}, \citenamefont {Rana}, \citenamefont {Bera},\ and\ \citenamefont {Lewenstein}}]{distributed_coherence}%
  \BibitemOpen
  \bibfield  {author} {\bibinfo {author} {\bibfnamefont {A.}~\bibnamefont {Streltsov}}, \bibinfo {author} {\bibfnamefont {S.}~\bibnamefont {Rana}}, \bibinfo {author} {\bibfnamefont {M.~N.}\ \bibnamefont {Bera}},\ and\ \bibinfo {author} {\bibfnamefont {M.}~\bibnamefont {Lewenstein}},\ }\bibfield  {title} {\bibinfo {title} {Towards resource theory of coherence in distributed scenarios},\ }\bibfield  {journal} {\bibinfo  {journal} {Physical Review X}\ }\textbf {\bibinfo {volume} {7}},\ \href {https://doi.org/10.1103/physrevx.7.011024} {10.1103/physrevx.7.011024} (\bibinfo {year} {2017}{\natexlab{b}})\BibitemShut {NoStop}%
\bibitem [{\citenamefont {Streltsov}\ \emph {et~al.}(2015)\citenamefont {Streltsov}, \citenamefont {Singh}, \citenamefont {Dhar}, \citenamefont {Bera},\ and\ \citenamefont {Adesso}}]{Streltsov2015}%
  \BibitemOpen
  \bibfield  {author} {\bibinfo {author} {\bibfnamefont {A.}~\bibnamefont {Streltsov}}, \bibinfo {author} {\bibfnamefont {U.}~\bibnamefont {Singh}}, \bibinfo {author} {\bibfnamefont {H.~S.}\ \bibnamefont {Dhar}}, \bibinfo {author} {\bibfnamefont {M.~N.}\ \bibnamefont {Bera}},\ and\ \bibinfo {author} {\bibfnamefont {G.}~\bibnamefont {Adesso}},\ }\bibfield  {title} {\bibinfo {title} {Measuring quantum coherence with entanglement},\ }\href {https://doi.org/10.1103/PhysRevLett.115.020403} {\bibfield  {journal} {\bibinfo  {journal} {Phys. Rev. Lett.}\ }\textbf {\bibinfo {volume} {115}},\ \bibinfo {pages} {020403} (\bibinfo {year} {2015})}\BibitemShut {NoStop}%
\bibitem [{\citenamefont {Killoran}\ \emph {et~al.}(2016)\citenamefont {Killoran}, \citenamefont {Steinhoff},\ and\ \citenamefont {Plenio}}]{Killoran2016_ConvertingCoherence}%
  \BibitemOpen
  \bibfield  {author} {\bibinfo {author} {\bibfnamefont {N.}~\bibnamefont {Killoran}}, \bibinfo {author} {\bibfnamefont {F.~E.~S.}\ \bibnamefont {Steinhoff}},\ and\ \bibinfo {author} {\bibfnamefont {M.~B.}\ \bibnamefont {Plenio}},\ }\bibfield  {title} {\bibinfo {title} {Converting nonclassicality into entanglement},\ }\href {https://doi.org/10.1103/PhysRevLett.116.080402} {\bibfield  {journal} {\bibinfo  {journal} {Phys. Rev. Lett.}\ }\textbf {\bibinfo {volume} {116}},\ \bibinfo {pages} {080402} (\bibinfo {year} {2016})}\BibitemShut {NoStop}%
\bibitem [{\citenamefont {Chitambar}\ and\ \citenamefont {Hsieh}(2016)}]{ChitambarHsieh2016}%
  \BibitemOpen
  \bibfield  {author} {\bibinfo {author} {\bibfnamefont {E.}~\bibnamefont {Chitambar}}\ and\ \bibinfo {author} {\bibfnamefont {M.-H.}\ \bibnamefont {Hsieh}},\ }\bibfield  {title} {\bibinfo {title} {Relating the resource theories of entanglement and quantum coherence},\ }\href {https://doi.org/10.1103/PhysRevLett.117.020402} {\bibfield  {journal} {\bibinfo  {journal} {Phys. Rev. Lett.}\ }\textbf {\bibinfo {volume} {117}},\ \bibinfo {pages} {020402} (\bibinfo {year} {2016})}\BibitemShut {NoStop}%
\bibitem [{\citenamefont {Zhu}\ \emph {et~al.}(2017)\citenamefont {Zhu}, \citenamefont {Ma}, \citenamefont {Cao}, \citenamefont {Fei},\ and\ \citenamefont {Vedral}}]{Ma2017}%
  \BibitemOpen
  \bibfield  {author} {\bibinfo {author} {\bibfnamefont {H.}~\bibnamefont {Zhu}}, \bibinfo {author} {\bibfnamefont {Z.}~\bibnamefont {Ma}}, \bibinfo {author} {\bibfnamefont {Z.}~\bibnamefont {Cao}}, \bibinfo {author} {\bibfnamefont {S.-M.}\ \bibnamefont {Fei}},\ and\ \bibinfo {author} {\bibfnamefont {V.}~\bibnamefont {Vedral}},\ }\bibfield  {title} {\bibinfo {title} {Operational one-to-one mapping between coherence and entanglement measures},\ }\href {https://doi.org/10.1103/PhysRevA.96.032316} {\bibfield  {journal} {\bibinfo  {journal} {Phys. Rev. A}\ }\textbf {\bibinfo {volume} {96}},\ \bibinfo {pages} {032316} (\bibinfo {year} {2017})}\BibitemShut {NoStop}%
\bibitem [{\citenamefont {Mekala}\ and\ \citenamefont {Sen}(2021)}]{Sen2021}%
  \BibitemOpen
  \bibfield  {author} {\bibinfo {author} {\bibfnamefont {A.}~\bibnamefont {Mekala}}\ and\ \bibinfo {author} {\bibfnamefont {U.}~\bibnamefont {Sen}},\ }\bibfield  {title} {\bibinfo {title} {All entangled states are quantum coherent with locally distinguishable bases},\ }\href {https://doi.org/10.1103/PhysRevA.104.L050402} {\bibfield  {journal} {\bibinfo  {journal} {Phys. Rev. A}\ }\textbf {\bibinfo {volume} {104}},\ \bibinfo {pages} {L050402} (\bibinfo {year} {2021})}\BibitemShut {NoStop}%
\bibitem [{\citenamefont {Cieśliński}\ \emph {et~al.}(2025)\citenamefont {Cieśliński}, \citenamefont {Vértesi}, \citenamefont {Kowalczyk},\ and\ \citenamefont {Laskowski}}]{Cieśliński_2025}%
  \BibitemOpen
  \bibfield  {author} {\bibinfo {author} {\bibfnamefont {P.}~\bibnamefont {Cieśliński}}, \bibinfo {author} {\bibfnamefont {T.}~\bibnamefont {Vértesi}}, \bibinfo {author} {\bibfnamefont {M.}~\bibnamefont {Kowalczyk}},\ and\ \bibinfo {author} {\bibfnamefont {W.}~\bibnamefont {Laskowski}},\ }\bibfield  {title} {\bibinfo {title} {How likely are you to observe non-locality with imperfect detection efficiency and random measurement settings?},\ }\href {https://doi.org/10.1088/1367-2630/ade90f} {\bibfield  {journal} {\bibinfo  {journal} {New Journal of Physics}\ }\textbf {\bibinfo {volume} {27}},\ \bibinfo {pages} {074504} (\bibinfo {year} {2025})}\BibitemShut {NoStop}%
\bibitem [{\citenamefont {Zyczkowski}\ and\ \citenamefont {Sommers}(2001)}]{Harr_Measures}%
  \BibitemOpen
  \bibfield  {author} {\bibinfo {author} {\bibfnamefont {K.}~\bibnamefont {Zyczkowski}}\ and\ \bibinfo {author} {\bibfnamefont {H.-J.}\ \bibnamefont {Sommers}},\ }\bibfield  {title} {\bibinfo {title} {Induced measures in the space of mixed quantum states},\ }\href {https://doi.org/10.1088/0305-4470/34/35/335} {\bibfield  {journal} {\bibinfo  {journal} {Journal of Physics A: Mathematical and General}\ }\textbf {\bibinfo {volume} {34}},\ \bibinfo {pages} {7111} (\bibinfo {year} {2001})}\BibitemShut {NoStop}%
\bibitem [{\citenamefont {{\ifmmode \dot{Z}\else \.{Z}\fi{}}yczkowski}\ and\ \citenamefont {Sommers}(2005)}]{random_states}%
  \BibitemOpen
  \bibfield  {author} {\bibinfo {author} {\bibfnamefont {K.}~\bibnamefont {{\ifmmode \dot{Z}\else \.{Z}\fi{}}yczkowski}}\ and\ \bibinfo {author} {\bibfnamefont {H.-J.}\ \bibnamefont {Sommers}},\ }\bibfield  {title} {\bibinfo {title} {Average fidelity between random quantum states},\ }\href {https://doi.org/10.1103/PhysRevA.71.032313} {\bibfield  {journal} {\bibinfo  {journal} {Phys. Rev. A}\ }\textbf {\bibinfo {volume} {71}},\ \bibinfo {pages} {032313} (\bibinfo {year} {2005})}\BibitemShut {NoStop}%
\bibitem [{\citenamefont {Chitambar}(2018)}]{DIO_reversibility}%
  \BibitemOpen
  \bibfield  {author} {\bibinfo {author} {\bibfnamefont {E.}~\bibnamefont {Chitambar}},\ }\bibfield  {title} {\bibinfo {title} {Dephasing-covariant operations enable asymptotic reversibility of quantum resources},\ }\href {https://doi.org/10.1103/PhysRevA.97.050301} {\bibfield  {journal} {\bibinfo  {journal} {Phys. Rev. A}\ }\textbf {\bibinfo {volume} {97}},\ \bibinfo {pages} {050301(R)} (\bibinfo {year} {2018})}\BibitemShut {NoStop}%
\bibitem [{\citenamefont {Zhao}\ \emph {et~al.}(2018)\citenamefont {Zhao}, \citenamefont {Liu}, \citenamefont {Yuan}, \citenamefont {Chitambar},\ and\ \citenamefont {Ma}}]{one-shot-dilution}%
  \BibitemOpen
  \bibfield  {author} {\bibinfo {author} {\bibfnamefont {Q.}~\bibnamefont {Zhao}}, \bibinfo {author} {\bibfnamefont {Y.}~\bibnamefont {Liu}}, \bibinfo {author} {\bibfnamefont {X.}~\bibnamefont {Yuan}}, \bibinfo {author} {\bibfnamefont {E.}~\bibnamefont {Chitambar}},\ and\ \bibinfo {author} {\bibfnamefont {X.}~\bibnamefont {Ma}},\ }\bibfield  {title} {\bibinfo {title} {One-shot coherence dilution},\ }\href {https://doi.org/10.1103/PhysRevLett.120.070403} {\bibfield  {journal} {\bibinfo  {journal} {Phys. Rev. Lett.}\ }\textbf {\bibinfo {volume} {120}},\ \bibinfo {pages} {070403} (\bibinfo {year} {2018})}\BibitemShut {NoStop}%
\bibitem [{\citenamefont {Ben~Dana}\ \emph {et~al.}(2017)\citenamefont {Ben~Dana}, \citenamefont {Garc{\'{\i}}a~D{\'{\i}}az}, \citenamefont {Mejatty},\ and\ \citenamefont {Winter}}]{channel_coherence}%
  \BibitemOpen
  \bibfield  {author} {\bibinfo {author} {\bibfnamefont {K.}~\bibnamefont {Ben~Dana}}, \bibinfo {author} {\bibfnamefont {M.}~\bibnamefont {Garc{\'{\i}}a~D{\'{\i}}az}}, \bibinfo {author} {\bibfnamefont {M.}~\bibnamefont {Mejatty}},\ and\ \bibinfo {author} {\bibfnamefont {A.}~\bibnamefont {Winter}},\ }\bibfield  {title} {\bibinfo {title} {Resource theory of coherence: Beyond states},\ }\href {https://doi.org/10.1103/PhysRevA.95.062327} {\bibfield  {journal} {\bibinfo  {journal} {Phys. Rev. A}\ }\textbf {\bibinfo {volume} {95}},\ \bibinfo {pages} {062327} (\bibinfo {year} {2017})}\BibitemShut {NoStop}%
\end{thebibliography}%

\appendix

\section{Phase Optimization of the Direct Scheme}
\label{app:direct_phase_optimization}

We derive the phase-optimized coherence yield of the Direct Scheme by determining the range of probabilities accessible through Alice's incoherent phase operations. Recall that the phase-dependent contribution to the yield is given by
\begin{equation}
    \mathcal{Y}_{\mathcal{P}^{\mathrm{dir}}}(\Omega)
    =
    h(p(\Omega))+h(\cos^2\theta),
\end{equation}
where $\Omega=(\gamma'-\delta)/2$ and
\begin{equation}
    p(\Omega)
    =
    \cos^2(\phi-\theta)
    -
    \sin(2\phi)\sin(2\theta)\sin^2\Omega.
\end{equation}

For $\phi,\theta\in[0,\pi/2]$, the function $p(\Omega)$ is continuous and, as $\sin^2\Omega$ varies over $[0,1]$, attains every value in the interval
\begin{equation}
    p\in
    \left[
    \cos^2(\phi+\theta),
    \cos^2(\phi-\theta)
    \right].
\end{equation}
Since the binary entropy $h(p)$ is maximized at $p=1/2$, the phase optimization separates naturally into two regimes.

If $1/2$ belongs to the accessible interval, Alice can choose a phase $\Omega^*$ such that $p(\Omega^*)=1/2$, yielding
\begin{equation}
    \mathcal{Y}_{\mathcal{P}^{\mathrm{dir}}}^*
    =
    1+h(\cos^2\theta).
\end{equation}
For nonzero $\sin(2\phi)\sin(2\theta)$, this condition is equivalent to the existence of a solution
\begin{equation}
    \sin^2\Omega^*
    =
    \frac{\cos(2(\phi-\theta))}
    {2\sin(2\phi)\sin(2\theta)}
    \in[0,1].
\end{equation}

Otherwise, the accessible interval lies entirely on one side of $p=1/2$. The global optimum follows directly from the monotonicity of binary entropy on either side of $p=1/2$ as
\begin{equation}
\begin{aligned}
    \mathcal{Y}_{\mathcal{P}^{\mathrm{dir}}}^*
    ={}&
    h(\cos^2\theta)\\
    &+\max\left\{
    h\big(\cos^2(\phi-\theta)\big),
    h\big(\cos^2(\phi+\theta)\big)
    \right\}.
\end{aligned}
\end{equation}

\section{Phase Optimization of the Entangled Scheme}
\label{app:entangled_phase_optimization}

In this appendix, we derive the stationary conditions for the Entangled Scheme yield $\mathcal{Y}_{\mathcal{P}^{\mathrm{ent}}}(\Omega) = H(\{p_{ij}(\Omega)\})$ with respect to the relative phase $\Omega = \gamma' - 2\delta$.

Using the shorthand notation $B = \frac{1}{4}\sin(2\phi)\sin^2(2\theta)$, the derivatives of the four joint probabilities $p_{ij}(\Omega)$ with respect to $\Omega$ are given by:

\begin{equation}
    \frac{dp_{00}}{d\Omega} = \frac{dp_{11}}{d\Omega} = -B\sin\Omega, \quad \text{and} \quad \frac{dp_{01}}{d\Omega} = \frac{dp_{10}}{d\Omega} = B\sin\Omega.
\end{equation}

Using the normalization of the probabilities, the derivative of their Shannon entropy satisfies $\frac{d h(p(\Omega))}{d \Omega} = -\sum_i \frac{ d p_i(\Omega)}{d \Omega} \log p_i(\Omega)$. Thus, we obtain
\begin{equation}
    \frac{d\mathcal{Y}_{\mathcal{P}^{\mathrm{ent}}}}
    {d\Omega}
    =
    B\sin\Omega
    \log\left(
    \frac{p_{00}(\Omega)p_{11}(\Omega)}
    {p_{01}(\Omega)^2}
    \right).
\end{equation}

For $B>0$, an interior stationary point $\Omega^*\in(0,\pi)$ must therefore satisfy $p_{00}(\Omega^*)p_{11}(\Omega^*)=p_{01}(\Omega^*)^2$. Expanding the probabilities yields
\begin{equation}
\begin{aligned}
    p_{00}p_{11}-p_{01}^2
    ={}&
    \frac14\sin^2(2\phi)\cos^2(2\theta)\\
    &+B\cos\Omega.
\end{aligned}
\label{eq:ent_probability_difference}
\end{equation}
Consequently, the interior stationary point is determined by
\begin{equation}
    \cos\Omega^*
    =
    -\frac{\sin(2\phi)\cos^2(2\theta)}
    {\sin^2(2\theta)}
    =
    -\frac{\sin(2\phi)}{\tan^2(2\theta)}.
\label{eq:ent_interior_phase}
\end{equation}

At $\theta=\pi/4$, the stationary condition is understood by continuity, yielding $\cos\Omega^*=0$. An interior solution exists when $\tan^2(2\theta)>\sin(2\phi)$, with equality corresponding to a boundary solution.

To establish global optimality, observe that the right-hand side of Eq.~(\ref{eq:ent_probability_difference}) is strictly decreasing in $\Omega\in(0,\pi)$ whenever $B>0$. Moreover, the logarithmic factor in the first derivative has the same sign as $p_{00}p_{11}-p_{01}^2$, while $B\sin\Omega>0$ throughout this interval. Thus, when no interior solution exists, the logarithmic factor remains positive throughout \((0,\pi)\), implying that the yield increases monotonically and attains its maximum at \(\Omega=\pi\). Conversely, when an interior solution $\Omega^*$ exists, the derivative changes sign from positive to negative at $\Omega^*$. Consequently, the yield increases before $\Omega^*$ and decreases thereafter, establishing that the interior solution is the global maximum and strictly dominates both boundary yields.

Since the probabilities depend on $\Omega$ only through $\cos\Omega$, we have $\mathcal{Y}_{\mathcal{P}^{\mathrm{ent}}}(2\pi-\Omega)=\mathcal{Y}_{\mathcal{P}^{\mathrm{ent}}}(\Omega)$. Therefore, restricting the optimization to $[0,\pi]$ entails no loss of generality. The degenerate case $B=0$ is independent of $\Omega$.

\section{Asymptotic Single-Qubit Optimizations}
\label{app:asymptotic_optimization}

We provide the details supporting the analytical pure-state optimization $F_\theta(C)$ and the numerical evaluation of the general single-qubit optimization $G_\theta(C)$ introduced in Sec.~\ref{sec:asymptotic}.

\subsection{Pure-State Optimization}

Consider a pure target state $\ket{\phi,\gamma'}=\cos\phi\ket{0}+e^{i\gamma'}\sin\phi\ket{1}$. By symmetry, we may restrict its tilt to $\phi\in[0,\pi/4]$, where Alice's coherence $h(\cos^2\phi)$ increases monotonically with $\phi$.

We first establish the minimum coherence cost $c_*(\theta)$ required to achieve unit yield. Let $\mathbf{a}$ and $\mathbf{b}$ denote Alice's and Bob's incoherent Bloch-sphere axes, respectively, with $\mathbf{a}\cdot\mathbf{b}=\cos(2\theta)$. A pure state with Bloch vector $\mathbf{r}$ is maximally coherent in Bob's basis precisely when $\mathbf{r}\cdot\mathbf{b}=0$. Among such states, Alice's coherence cost is minimized by maximizing $|\mathbf{r}\cdot\mathbf{a}|$, giving
\begin{equation}
\begin{aligned}
\max_{\substack{|\mathbf{r}|=1\\
\mathbf{r}\cdot\mathbf{b}=0}}
|\mathbf{r}\cdot\mathbf{a}|
&=\sqrt{1-(\mathbf{a}\cdot\mathbf{b})^2}\\
&=|\sin(2\theta)|.
\end{aligned}
\end{equation}
Consequently, the minimum Alice-coherence cost of a maximally Bob-coherent pure state is
\begin{equation}
    c_*(\theta)
    =h\left(\frac{1+|\sin(2\theta)|}{2}\right),
\end{equation}
as stated in Proposition~\ref{prop:asymptotic_pure}.

It remains to determine the optimal yield when Alice's coherence budget is insufficient to reach this threshold. As established in Appendix~\ref{app:direct_phase_optimization}, optimization over the preparation phase allows the probability of Bob's first basis outcome to attain any value in
\begin{equation}
    p\in
    [\cos^2(\phi+\theta),\cos^2(\phi-\theta)].
\end{equation}
For $C<c_*(\theta)$, this interval does not contain $1/2$ for any admissible $\phi$. When $\theta<\pi/4$, the interval lies above $1/2$, and its lower endpoint decreases towards $1/2$ as $\phi$ increases. Conversely, for $\theta>\pi/4$, the interval lies below $1/2$, and its upper endpoint increases towards $1/2$.

Since binary entropy increases as its argument approaches $1/2$, the phase-optimized yield is monotonically increasing with $\phi$ throughout the admissible range. Thus, the optimal pure-state target exhausts Alice's coherence budget, with
\begin{equation}
    h(\cos^2\phi_C)=C,
\end{equation}
establishing the expression for $F_\theta(C)$ given in Eq.~\eqref{eq:asymptotic_combined_bound}.

\subsection{General Single-Qubit Optimization}

We now consider arbitrary single-qubit target states
\begin{equation}
    \sigma=\frac12(I+\mathbf r\cdot\boldsymbol{\sigma}),
\end{equation}
where $R=|\mathbf r|\leq1$ is the Bloch-vector length and $z=|r_z|$ is the magnitude of its component along Alice's incoherent axis.

Introducing $H(x)=h((1+x)/2)$, the relative entropy of coherence in Alice's basis is
\begin{equation}
    C_A(\sigma)=H(z)-H(R).
\end{equation}
Similarly, writing $\mathbf b$ for Bob's incoherent axis, we obtain
\begin{equation}
    C_B(\sigma)=H(|\mathbf r\cdot\mathbf b|)-H(R).
\end{equation}

For fixed $R$ and $z$, optimizing the azimuthal orientation of $\mathbf r$ minimizes its absolute projection onto Bob's axis, giving
\begin{equation}
    \min_{\gamma'}
    |\mathbf r\cdot\mathbf b|
    =
    \left[
    z|\cos(2\theta)|
    -
    \sqrt{R^2-z^2}|\sin(2\theta)|
    \right]_+,
\end{equation}
where $[x]_+=\max\{0,x\}$.

Consequently, the evaluation of $G_\theta(C)$ reduces to maximizing
\begin{equation}
\begin{aligned}
    C_B^{\mathrm{opt}}(R,z)
    ={}&
    H\left(
    \left[
    z|\cos(2\theta)|
    -
    \sqrt{R^2-z^2}|\sin(2\theta)|
    \right]_+
    \right)
    \\& -H(R)
\end{aligned}
\end{equation}
subject to
\begin{equation}
    0\leq z\leq R\leq1,
    \qquad H(z)-H(R)\leq C.
\end{equation}

For numerical evaluation, we first optimize over the admissible Bloch-vector parameters at fixed target coherence $C'$, and subsequently maximize the resulting yield over $0\leq C'\leq C$ to obtain $G_\theta(C)$.

The numerical investigation indicates that, similar to the pure state strategy, allowing \(C'<C\) does not improve the optimal yield throughout the sampled parameter range. However, mixed target states can yield a small improvement over the analytically optimized pure-state construction, as illustrated in Fig.~\ref{fig:asymptotic_ratio}.

\end{document}